\documentclass[aps,pra,twocolumn,floats,amsmath,amssymb,superscriptaddress]{revtex4-1}
\usepackage[utf8]{inputenc}
\usepackage{graphicx}
\usepackage{epsfig}
\usepackage{amsfonts}
\usepackage{amsmath}
\usepackage{natbib}
\usepackage{multirow}
\usepackage{bm}
\usepackage{epstopdf}
\usepackage{ulem}
\usepackage{color}
\usepackage{hyperref}
\usepackage{mathtools}
\usepackage{cases}
\usepackage{physics}
\usepackage{amssymb}
\usepackage[noabbrev]{cleveref}
\newcommand{\ie}{{\it i.e.}, }
\newcommand{\eg}{{\it e.g.}, }

\newcommand{\exciting}{{\usefont{T1}{lmtt}{b}{n}exciting}}
\Crefname{equation}{Eq.}{Eqs.}
\def\bq{{\bf q}}
\def\bk{{\bf k}}
\def\bG{{\bf G}}
\def\br{{\bf r}}
\def\bR{{\bf R}}
\def\GW{{$G_0W_0$}}
\usepackage{xcolor}
\begin{document}
\title{Efficient \GW\ and BSE calculations of heterostructures within an all-electron framework}

\author{Maximilian Schebek$^\dagger$}
\thanks{These authors contributed equally to this work.}
\affiliation{Physics Department and CSMB, Humboldt-Universit\"{a}t zu Berlin, zum Großen Windkanal 2, 12489 Berlin, Germany}
\affiliation{Fachbereich Physik, Freie Universit\"{a}t Berlin, D-14195 Berlin, Germany}

\author{Ignacio \surname{Gonzalez Oliva}}
\thanks{These authors contributed equally to this work.}
\affiliation{Physics Department and CSMB, Humboldt-Universit\"{a}t zu Berlin, zum Großen Windkanal 2, 12489 Berlin, Germany}

\author{Claudia Draxl}
\affiliation{Physics Department and CSMB, Humboldt-Universit\"{a}t zu Berlin, zum Großen Windkanal 2, 12489 Berlin, Germany}
\affiliation{European Theoretical Spectroscopic Facility (ETSF)}

\begin{abstract}
The combination of two-dimensional materials into heterostructures offers new opportunities for the design of optoelectronic devices with tunable properties. However, computing electronic and optical properties of such systems using state-of-the-art methodology is challenging due to their large unit cells. This is in particular so for highly-precise all-electron calculations within the framework of many-body perturbation theory, which come with high computational costs. Here, we extend an approach that allows for the efficient calculation of the non-interacting polarizability, previously developed for planewave basis sets, to the (linearized) augmented planewave (L)APW method. This approach is based on an additive ansatz, which computes and superposes the polarizabilities of the individual components in their respective unit cells. We implement this formalism in the \GW\ module of the \exciting\ code and implement an analogous approach for BSE calculations. This allows the calculation of highly-precise optical spectra at low cost. So-obtained results of the quasi-particle band structure and optical spectra are demonstrated for bilayer WSe$_2$ and pyridine@MoS$_2$ in comparison with exact reference calculations. \\

\noindent
$^\dagger$Corresponding author: m.schebek@fu-berlin.de 
\end{abstract}

\maketitle

\section{Introduction}
The combination of two-dimensional (2D) materials into weakly bound van der Waals (vdW) heterostructures has received increasing attention in the last decade due to the possibility of tailoring their electronic and optical properties towards desired features and functions  \cite{Wang2021_review}. A class of materials with promising opto-electronic properties are transition metal dichalcogenides (TMDCs)~\cite{Molina2015,Wang2018,Devakul2021-gi,Caruso2022}, which can be combined into multi-layered structures. Particularly interesting with a high potential for new optoelectronic and photonic devices is the combination of TMDCs with organic molecules \cite{Ji2022}. Such hybrid inorganic organic systems (HIOS) allow to enhance the specific strengths of the individual materials while compensating for their weaknesses \cite{Draxl2014,Hofmann2021,GonzalezOliva2023,Liu2025,Benson2024}. 

For an efficient exploration of possible combinations of materials, accurate theoretical and computational methods are essential \cite{Draxl2014, Hofmann2021, GonzalezOliva2023, Liu2025}. In particular, a reliable description of the band structure and the resulting energy-level alignment \cite{Xuan2019,Liu2019,Adeniran2021,Frimpong2024} of such an interface is a must, as it is the basis for understanding the material's excitation spectra \cite{Palummo2019,Wang2020,Wang2021,GonzalezOliva2022}, which in turn determines whether it is suitable for a given application. Approaches based solely on density-functional theory (DFT) are not useful for vdW heterostructures and HIOS, since many-body effects not only govern their response to light but also their electronic structure. For example, the wrong type of energy-level alignment of HIOS has been predicted not only by semi-local DFT but also by hybrid functionals~\cite{GonzalezOliva2023}. For these reasons, many-body perturbation theory (MBPT), in particular the \GW~approximation and the Bethe-Salpeter equation (BSE), are the state-of-the-art for studying electronic and optical properties of such materials. 

Unfortunately, both \GW\ and BSE calculations are computationally very demanding and therefore often not feasible for large unit cells. This is the case for interface systems, which typically require combining supercells of the pristine constituents. One of the major bottlenecks in both methods is the calculation of the polarizability, which scales as $O(N^4)$ with the number of atoms $N$ in the unit cell \cite{Hybertsen1985,Hybertsen1986}. Therefore, recent efforts have focused on reducing the computational cost of polarizability calculations, either by developing algorithms with a more favorable scaling \cite{liu_cubic_gw, Duchemin2021-cv, Neuhauser2014} or by employing efficient approximations. A particularly efficient approximation for weakly-bound systems consists in an additive ansatz, where the polarizability of the heterostructure is obtained by summing the contributions of the individual components \cite{Xuan2019,Liu2019}. This comes with the additional advantage that the polarizability of the supercell of a constituent does not need to be computed for that supercell but can be obtained from the respective unit cell by a folding (expansion) procedure. The combination of these two approaches, called {\it expansion} \& {\it addition screening} (EAS) here, has been successfully applied to the calculation of the energy-level alignment at molecule-metal interfaces~\cite{Xuan2019,Liu2019}. However, in these works, a planewave representation of the polarizability has been used, and thus it is not directly applicable to other basis sets, in particular those used in all-electron codes \cite{Jiang2013,Nabok2016,Ren2012}. However, all-electron calculations are very important for reference calculations, \eg for benchmarking new implementations or for assessing the impact of pseudopotentials \cite{Lejaeghere2016,Azizi2025}. 

In this work, we extend the EAS approach to the so-called mixed-product basis used in excited-state calculations based on the (linearized) augmented planewave (L)APW method \cite{Singh2005}. We derive the formalism and implement it within the \GW~module of the \exciting\ code \cite{Gulans2014, Nabok2016}. Furthermore, we implement an analogous method for the BSE framework \cite{Vorwerk2019}. These developments allow for the calculation of the electronic structure as well as of the optical absorption spectra of large interfaces with high accuracy at significantly reduced computational cost compared to standard approaches. We demonstrate this by computing the quasi-particle (QP) band structures and optical spectra of two prototypical systems, namely the WSe$_2$ bilayer and the HIOS pyridine@MoS$_2$. 

The manuscript is structured as follows. The theoretical background covering the planewave-based EAS method is revisited in Section~\ref{section:theoretical_background}. The mixed-product basis is summarized in Section~\ref{section:mixed_product_basis}. In Sections~\ref{section:expansion_step} and~\ref{section:addition_step}, we describe the expansion and addition steps within our basis. In Section~\ref{section:results}, we present the so obtained QP band structure and optical spectra of the selected systems and compare them with the exact calculations.

\section{Background}

\subsection{Polarizability in the EAS method with a planewave basis}
\label{section:theoretical_background}

The screened Coulomb potential $W$ is central to both $GW$ and BSE calculations, since screening is essential for both the electron-electron and the electron-hole interaction, respectively\cite{Rohlfing2000, Hybertsen1986}. It is defined as
\begin{align}\label{eq:W}
    W(\mathbf{r}, \mathbf{r}'; \omega) = \int \! \text{d}\mathbf{r}_1 \, \epsilon^{-1} (\mathbf{r}, \mathbf{r}_1; \omega) \, v(\mathbf{r}_1, \mathbf{r}')  \:, 
\end{align}
where $v(\mathbf{r}, \mathbf{r}') = |\mathbf{r} - \mathbf{r}'|^{-1}$  is the bare Coulomb interaction. $\epsilon$ denotes the dielectric function which, in a symbolic notation, can be obtained from the non-interacting random-phase approximation (RPA) polarizability $P$ as $\epsilon=1 - vP$. $P$ is typically expanded into a set of basis functions \{$\chi^{\bq}_i$(\br)\} according to 
\begin{equation}\label{eq:pol_expansion}
    P(\br,\br', \omega) = \sum_{\bq}^{\rm BZ}\sum_{ij} P_{ij}(\bq, \omega) \,  \chi_i^\bq(\br) \, \Big[  \chi_j^\bq(\br') \Big]^* \:,
\end{equation}
where the matrix elements can be obtained as
\begin{align}  
    P_{ij}(\bq, \omega) & = \int\!\!\int \text{d}\br \, \text{d}\br' \, \Big[  \chi_i^\bq(\br) \Big]^* \, P(\br,\br', \omega) \, \chi_j^\bq(\br') \:, \label{eq:pol_matrix_elements} 
\end{align}
In this work, we consider two different basis sets for the expansion of the polarizability in Eq.~\ref{eq:pol_expansion}, namely a mixed-product basis for $GW$ calculations and planewaves for the BSE. From this point on, we omit the frequency dependence of the polarizability for simplicity since what follows applies to all frequencies. 

The key idea of the additive ansatz for the polarizability \cite{Xuan2019, Liu2019}, \ie $P=P^1+P^2$, is to obtain the contributions from the individual systems only, thus neglecting effects coming from the interface. This additive approach is applicable to systems with negligible covalent interactions, \eg heterostructures that can be easily separated into two distinct components. Using a planewave expansion of the polarizability in terms of the reciprocal lattice vectors $\bG$, \ie $\chi^\bq_i(\br) \rightarrow \chi^\bq_\bG(\br) \propto e^{\text{i}(\bq+\bG)\br}$, this approach can be easily implemented to obtain the coefficients $P^\mu_{\bG\bG'}(\bq)$ and the subsequent addition according to
\begin{align}\label{eq:add_polarizability}
P_{\bG\bG'}(\bq) = P^1_{\bG\bG'}(\bq) + P^2_{\bG\bG'}(\bq) \:.
\end{align}
Due to the quartic scaling of the polarizability, the additive approach itself already offers significant computational savings. Further savings can be achieved when expanding the individual polarizabilities. In the expansion step, the smallest possible unit cell is identified for each individual component for computing the polarizability. Then the polarizability of the respective component in the supercell of the heterostructure is obtained by an expansion (folding) procedure. This expansion requires to map  all $(\bq_{\text{uc}}+\bG_{\text{uc}})$ vectors of the unit cell to the corresponding $(\bq+\bG)$ vectors of the supercell:
\begin{align}\label{eq:pol_map_uc_sc}
  P^\mu_{\bG_{\rm uc}\bG'_{\rm uc}}(\bq_{\rm uc}) \rightarrow  P^\mu_{\bG\bG'}(\bq) \:,
\end{align}
with $\bq_{\text{uc}}+\bG_{\text{uc}} = \bq+\bG $ and $\bq_{\text{uc}}+\bG'_{\text{uc}} = \bq+\bG' $.  The expansion is therefore only possible for commensurate grids $ \{\bq + \bG\} \subseteq\{\bq_{\rm uc} + \bG_{\rm uc}\} $.
 
\subsection{The mixed product basis}
\label{section:mixed_product_basis}

We extend the EAS approach in the framework of the (L)APW method, which employs augmented planewaves and local orbitals (lo) as basis functions. In this method, the unit cell is partitioned into muffin-tin (MT) spheres and the interstitial region (I). The (L)APW part of the basis is shown in Eqs.~\ref{eq:lapw_mt} and  \ref{eq:lapw_int}. More details are presented in Appendix~\ref{appendix:mixed_basis}. 

The \GW\ implementation in the all-electron full-potential code \exciting\ employs an auxiliary basis~\cite{Friedrich2010, Jiang2013, Nabok2016} that provides a highly flexible representation of the polarizability. This so-called mixed-product basis, used to optimally represent the product of two Kohn-Sham states \cite{Nabok2016}, is defined as: 
\begin{widetext}
\begin{numcases}{\chi^{\bq}_j (\br) =}
    \gamma^\bq_{\alpha N L M} (\mathbf{r}^\alpha)= e^{\text{i}\bq \cdot \br_{\alpha}} \, v_{\alpha NL} (r^{\alpha}) \, Y_{LM} (\hat{\br}^{\alpha}) & $r^{\alpha} < R^{\alpha}_{\text{{\rm MT}}}$ \:, \label{eq:mt_mpb} \\[5pt]
    \Lambda_i^\bq (\br) = \frac{1}{\sqrt{\Omega}} \sum_{\bG} \Tilde{S}^{*}_{\bG i} \, e^{\text{i}(\bG + \bq) \cdot \br} \, \theta_{I} (\br) & $\br \in \text{I}$ \:. \label{eq:int_mpb}
\end{numcases}
\end{widetext}
where $R^{\alpha}_{\text{{\rm MT}}}$ is the radius of the sphere of atom $\alpha$, centered at position $\bm{\tau}_{\alpha}$, and the difference vector from this center is labeled as $\mathbf{r}^\alpha = \mathbf{r} - \bm{\tau}_\alpha$. The unit-cell volume is denoted by $\Omega$. Inside the {\rm MT} region, the basis is constructed in terms of radial functions $v_{\alpha NL} (r^{\alpha})$ for an atom of index $\alpha$, and  spherical harmonics  $Y_{LM} (\hat{\br}^{\alpha})$ with quantum numbers $L$ and $M$ (Eq.~\ref{eq:mt_mpb}). In the interstitial region, a linear combination of planewaves is used, where $\Tilde{S}_{\bG i}$ are the renormalized planewave eigenvectors ensuring orthogonality and $\theta_{I} (\br)$ is the step function, which is $0$ within the {\rm MT}s and $1$ outside (Eq.~\ref{eq:int_mpb}).
For more details on the \GW\ module the reader is referred to Ref. \cite{Nabok2016} and Appendix~\ref{appendix:mixed_basis}.

\section{The EAS approach in the mixed-product basis}
\subsection{The expansion step}
\label{section:expansion_step}

\begin{figure}[h!]
\begin{center}
\includegraphics[width=0.48\textwidth]{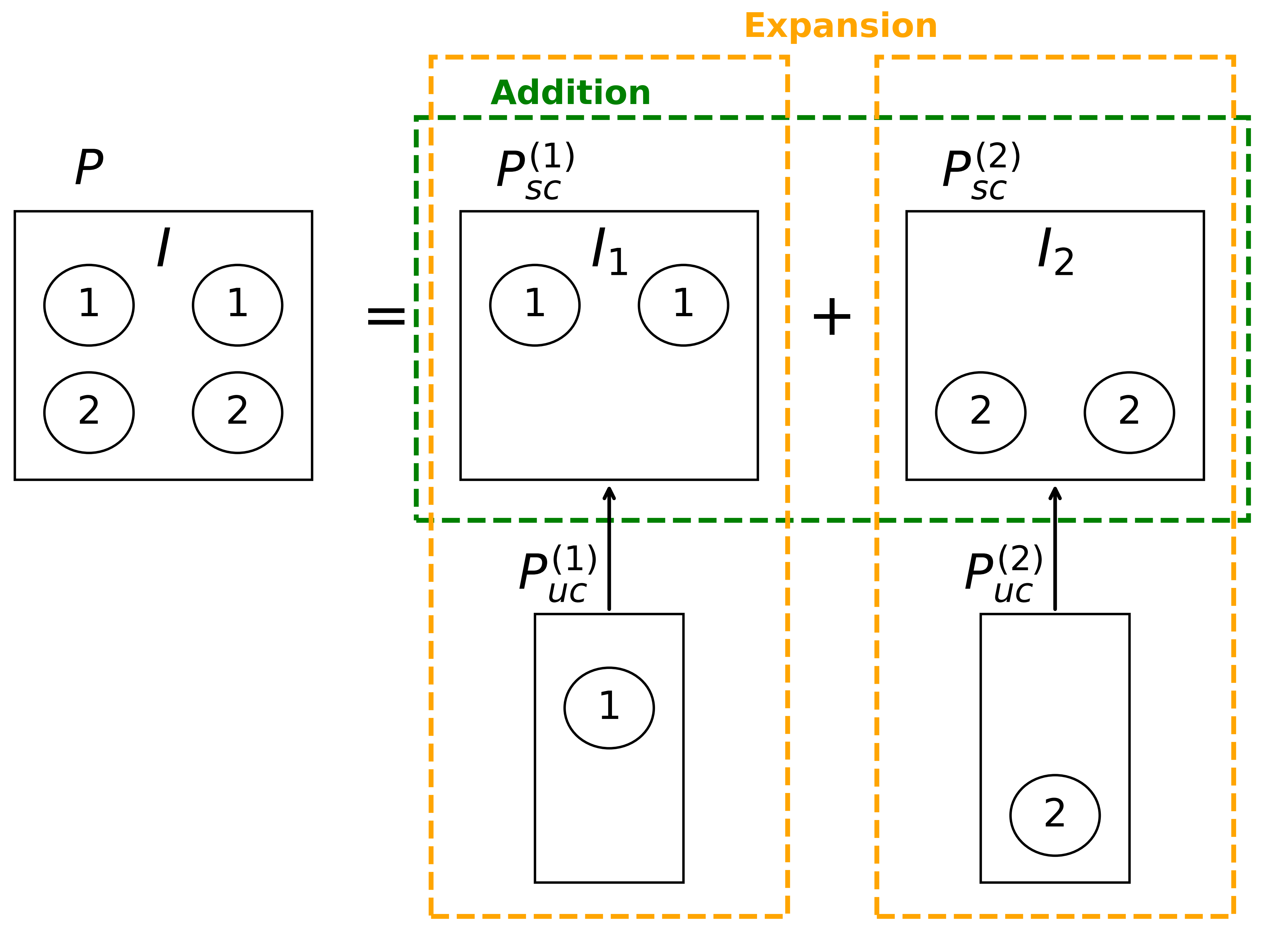}%
\caption{Scheme of the {\it expansion} \& {\it addition} method. The polarizability of the heterostructure, $P$, is approximated as the sum of the individual polarizabilities, $P_{\text{sc}}^{(1)}$ and $P_{\text{sc}}^{(2)}$, which are obtained as the expansions of the respective polarizabilities computed in their unit cells, $P_{\text{uc}}^{(1)}$ and $P_{\text{uc}}^{(2)}$, respectively.}
\label{fig:eas_scheme}
\end{center}
\end{figure}

In the expansion step, the polarizability of the supercell is obtained by expanding the polarizability of the smallest unit cell of each component through a folding procedure (orange box in Fig.~\ref{fig:eas_scheme}). To begin, it is convenient to partition the polarizability into 4 different blocks according to
\begin{align}
    P = 
\begin{pmatrix}
    P_{{\rm MT},{\rm MT}} & P_{{\rm MT},I} \\[5pt]
    P_{I,{\rm MT}} & P_{I,I}  \label{eq:general_pol_matrix}
\end{pmatrix} \:.
\end{align}
In this notation, the block $P_{XY}$ includes all matrix elements $P_{ij}$ (Eq.~\ref{eq:pol_matrix_elements}) with $\chi_i(\br)$ ($\chi_j(\br')$) being defined in the region $X$ ($Y$). Note that in the preceding equation, the index {\rm MT} does not refer to one specific muffin-tin sphere but includes all muffin-tin spheres of the system of interest. Due to the nature of the mixed-product basis, the {\rm MT} spheres and the interstitial region require very different approaches for performing the expansion. 

We first detail the procedure for the case in which both coordinates are in atomic spheres. The matrix elements of the supercell, $P^{\text{sc}}_{{\rm MT},{\rm MT}}$, can be obtained in terms of the matrix elements of the unit cell, $P^{\text{uc}}_{{\rm MT},{\rm MT}}$, via a transformation from real to reciprocal space as follows: 
\begin{align}
    P^{\text{sc}}_{\alpha_{\text{sc}} \beta_{\text{sc}}} (\bq^{\text{sc}}) & = \sum_{\mathbf{R}^{\text{sc}}}  e^{-\text{i} \bq^{\text{sc}} \cdot \mathbf{R}^{\text{sc}}} \, P^{\text{sc}}_{\alpha_{\text{sc}} \beta_{\text{sc}}} (\mathbf{R}^{\text{sc}}) \:, \nonumber \\[5pt]
    & = \frac{N^{\text{sc}}_{\bq}}{N_{\bq}^{\text{uc}}} \sum_{\bq^{\text{uc}}} e^{\text{i} \bq^{\text{uc}} \cdot (\mathbf{R}_{\beta}^{\text{uc}} - \mathbf{R}_{\alpha}^{\text{uc}})} \,  P^{\text{uc}}_{\alpha_{\text{uc}} \beta_{\text{uc}}} (\bq^{\text{uc}}) \:,
    \label{eq:expansion_mt}
\end{align}
where $\textbf{R}$ are the real-space lattice vectors, $N_{\bq}$ the number of \textbf{q}-points, and the subscript uc (sc) refers to unit-cell (supercell) quantities. The difference of real-space lattice vectors is rewritten in terms of the atomic positions $\tau$,
\begin{align}
     \bm{\tau}_{\alpha_{\text{sc}}} & = \bm{\tau}_{\alpha_{\text{uc}}} + \mathbf{R}_{\alpha}^{\text{uc}} \:, \label{eq:atomic_position}
\end{align}
as
\begin{align}   
    \mathbf{R}_{\beta}^{\text{uc}} - \mathbf{R}_{\alpha}^{\text{uc}} & = (\bm{\tau}_{\beta_{\text{sc}}} - \bm{\tau}_{\beta_{\text{uc}}}) - (\bm{\tau}_{\alpha_{\text{sc}}} - \bm{\tau}_{\alpha_{\text{uc}}}) \:. 
    \label{eq:phase_factor_mt}
\end{align}
Once the equivalent atoms in the supercell have been identified using Eq.~\ref{eq:atomic_position}, the implementation of Eq.~\ref{eq:expansion_mt} is straightforward. 

For the block only involving the interstitial region, $P_{I,I}$, where planewaves are employed, we apply the mapping defined in Eq.~\ref{eq:pol_map_uc_sc} to the interstitial mixed-product basis functions (see Eq.~\ref{eq:int_mpb}). This is achieved by first transforming the mixed-product basis to the original set of planewaves using
\begin{align}
    P_{\bG\bG'}(\bq) &  = \sum_{ij}P_{ij}(\bq) \, \langle \chi_i^\bq | \chi_\bG^{\bq} \rangle^* \, \langle \chi_j^{\bq} | \chi_{\bG'}^\bq \rangle \:,
    \label{eq:transformation_IPW}
\end{align}
where the overlap between these two bases is given by
\begin{align}
    \langle \chi_i^\bq | \chi_\bG^{\bq} \rangle = & \frac{1}{\Omega}\int \text{d}\br \, \sum_{\bG'}\tilde{S}_{\bG'i} \, e^{-\text{i}(\bq+\bG')\br} \, e^{\text{i}(\bq+\bG)\br} \:, \nonumber \\[5pt]
    = & \sqrt{\lambda_i}\, S_{\bG i}  \:.
    \label{eq:transformation_mixed_ipw}
\end{align}
Here, $\lambda$ are the eigenvalues of the planewave overlap matrix (see Eq.~\ref{eq:overlap_planewave_product}) and we have used Eq.~\ref{eq:planewave_eigenvector}. Second, the polarizability is expanded according to Eq.~\ref{eq:pol_map_uc_sc} to obtain it for the supercell in the planewave basis. Finally, we transform this matrix back to the mixed-product basis of the supercell using the inverse transformation of Eq.~\ref{eq:transformation_IPW}.

The polarizability matrix has two off-diagonal blocks, \ie $P_{{\rm MT}/I}$ and $P_{I/{\rm MT}}$. In these cases, the procedure described above for $P_{{\rm MT},{\rm MT}}$ ($P_{I,I}$) is applied only to the rows (columns) and columns (rows) of $P_{{\rm MT},I}$ and $P_{I,{\rm MT}}$, respectively. Once the expansion is performed, the polarizability of the supercell is added to its counterpart coming from the other component of the heterostructure as described in the next section. 

\subsection{The addition step}
\label{section:addition_step}

In the addition step, we construct the polarizability of the heterostructure by calculating the sum of the individual contributions according to $P^{\text{HS}}(\bq) = P^1(\bq) + P^2(\bq)$ (green box in Fig.~\ref{fig:eas_scheme}). For this part, we assume that the polarizabilities of the individual components are available, either by a direct calculation for the supercell or by applying the expansion procedure presented in the preceding section. This addition procedure is again not trivial due to the dual representation of our basis, as it requires to transform the polarizability from the bases of the individual components, $\{\chi^{\mu,\bq}_a\}$, to the basis of the heterostructure, $\{\chi^{\bq}_A\}$. 

For weakly bound systems, the coefficients of the individual polarizability of component $\mu$ in the basis of the full heterostructure can be approximated as 
\begin{align}
    P^\mu_{AB}(\bq) & =  \sum_{ab}C_{aA}^{\mu*}(\bq) \, P^\mu_{ab}(\bq) \, C^\mu_{bB}(\bq) \:,
    \label{eq:transform_final}
\end{align}
where the overlap matrix elements $C^\mu_{aA}(\bq)=\braket{\chi^{\mu,\bq}_a}{\chi_A^{\bq}}$ are introduced. Similar to Eq.~\ref{eq:general_pol_matrix}, it is convenient to partition the polarizability of the heterostructure into different blocks according to
\begin{align} \label{eq:pol_hs_add}
    P_{\text{HS}} =
\begin{pmatrix}
    P_{{\rm MT}_1,{\rm MT}_1} & P_{{\rm MT}_1,{\rm MT}_2} &  P_{{\rm MT}_1,I}\\[5pt]
    P_{{\rm MT}_2,{\rm MT}_1} & P_{{\rm MT}_2,{\rm MT}_2} &  P_{{\rm MT}_2,I}\\[5pt]
    P_{I,{\rm MT}_1} & P_{I,{\rm MT}_2} &  P_{I,I} \\
\end{pmatrix} \:,
\end{align}
where each term represents a block of the polarizability with the basis functions coming from the different regions. We take advantage of the fact that the {\rm MT} region of the heterostructure $\Omega_{\rm {\rm MT}}$ can be further divided into the two individual {\rm MT} regions, \ie $\Omega_{\rm {\rm MT}}=\Omega_{\rm {\rm MT}_1} \cup \Omega_{\rm {\rm MT}_2}$ with $\Omega_{\rm {\rm MT}_1} \cap \Omega_{\rm {\rm MT}_2}=0$.  For convenience, we omit the \textbf{q}-vector index from now on. 

We begin by constructing the matrix $C^{\mu}$ for component $\mu$, which can be partitioned according to 
\begin{equation}\label{eq:mat_overlap}
    C^\mu =
\begin{pmatrix}
    C_{{\rm MT}_\mu/{\rm MT}_1} & C_{{\rm MT}_\mu/{\rm MT}_2} & C_{{\rm MT}_\mu/I} \\[5pt]
    C_{I_\mu/{\rm MT}_1} & C_{I_\mu/{\rm MT}_2} & C_{I_\mu/I}
\end{pmatrix} \:,
\end{equation}
where $I_{\mu}$ is the interstitial region of system $\mu$.

The first term, $C_{{\rm MT}_\mu/{\rm MT}_\nu}$, corresponds to the overlap between basis functions in the {\rm MT} region of one of the individual layers and basis functions in the {\rm MT} region of layer $\nu$ in the heterostructure. Here, we assume that the radial functions in the spheres of the individual component and the heterostructure are the same, owing to the weak interaction between the two. Furthermore, since the two regions are disjunct, we can conclude that $C_{{\rm MT}_\mu/{\rm MT}_\nu} = \delta_{\mu\nu}$.

The next term that we consider is $C_{{\rm MT}_\mu/I}$. We note that the {\rm MT} region of any individual component does not overlap with the I region of the heterostructure, so $ C_{{\rm MT}_\mu/I}=0$. Similarly, for the terms $C_{I_\mu/{\rm MT}_1}$ and $C_{I_\mu/{\rm MT}_2}$, there is no overlap between the functions of the {\rm MT} and I regions of the same layer, hence $C_{I_\mu/{\rm MT}_\mu} =0$. 

The only nontrivial blocks are then $C_{I_\mu/{\rm MT}_\nu}$ and $C_{I_\mu/I}$, where an element of these matrices is given by 
\begin{align}\label{eq:C_non_zero}
    C^{\mu}_{aA} & =\sum_{\bG}\tilde{S}^{\mu}_{\bG a} \, w^{*}_{A \bG} \:,
\end{align}
with the distinction that for $C_{I_\mu/{\rm MT}_\nu}$ the index $a$ ($A$) corresponds to the functions of the I ({\rm MT}) region of the layer $\mu$ ($\nu$). For $C_{I_\mu/I}$ the index $a$ corresponds to the I region of the layer $\mu$, while $A$ corresponds to the I region of the heterostructure. In Eq.~\ref{eq:C_non_zero}, we have introduced $w$ as the overlap matrix elements between a mixed-basis function and a planewave (see Eq.~\ref{eq:overlap_integrals}). In Appendix~\ref{appendix:overlap_matrix}, we show the derivations of the expressions for the different blocks of $C^{\mu}$.

We now show the general expressions to calculate the components of Eq.~\ref{eq:pol_hs_add} according to Eq.~\ref{eq:transform_final}. Concretely, we obtain 
\begin{align}
P_{{\rm {\rm MT}}_\mu,{\rm {\rm MT}}_\mu} = & P^\mu_{{\rm {\rm MT}}_\mu,{\rm {\rm MT}}_\mu} + (C^\nu_{I_\nu,{\rm {\rm MT}}_\mu})^* P^\nu_{I_\nu,I_\nu}C^\nu_{I_\nu,{\rm {\rm MT}}_\mu} \:, \nonumber  \\[5pt]
P_{{\rm {\rm MT}}_\mu, {\rm {\rm MT}}_\nu} = & P^\mu_{{\rm {\rm MT}}_\mu,I_\mu}C^\mu_{I_\mu,{\rm {\rm MT}}_\nu} + (C^\nu_{I_\nu,{\rm {\rm MT}}_\mu})^*P^\nu_{I_\nu,{\rm {\rm MT}}_\nu} \:, \nonumber  \\[5pt]
P_{{\rm {\rm MT}}_\mu, I} = & P^\mu_{{\rm {\rm MT}}_\mu,I_\mu}C^\mu_{I_\mu,I} + (C^\nu_{I_\nu,{\rm {\rm MT}}_\mu})^*P^\nu_{I_\nu,I_\nu}C^\nu_{I_\nu,I} \nonumber \:, \\[5pt]
P_{I,I} = & \sum_\mu (C^\mu_{I_\mu,I})^* \, P^\mu_{I_\mu,I_\mu} \, C^\mu_{I_\mu,I}  \:.
\end{align}

\begin{figure*}[t!]
\begin{center}
\includegraphics[width= 0.8\textwidth]{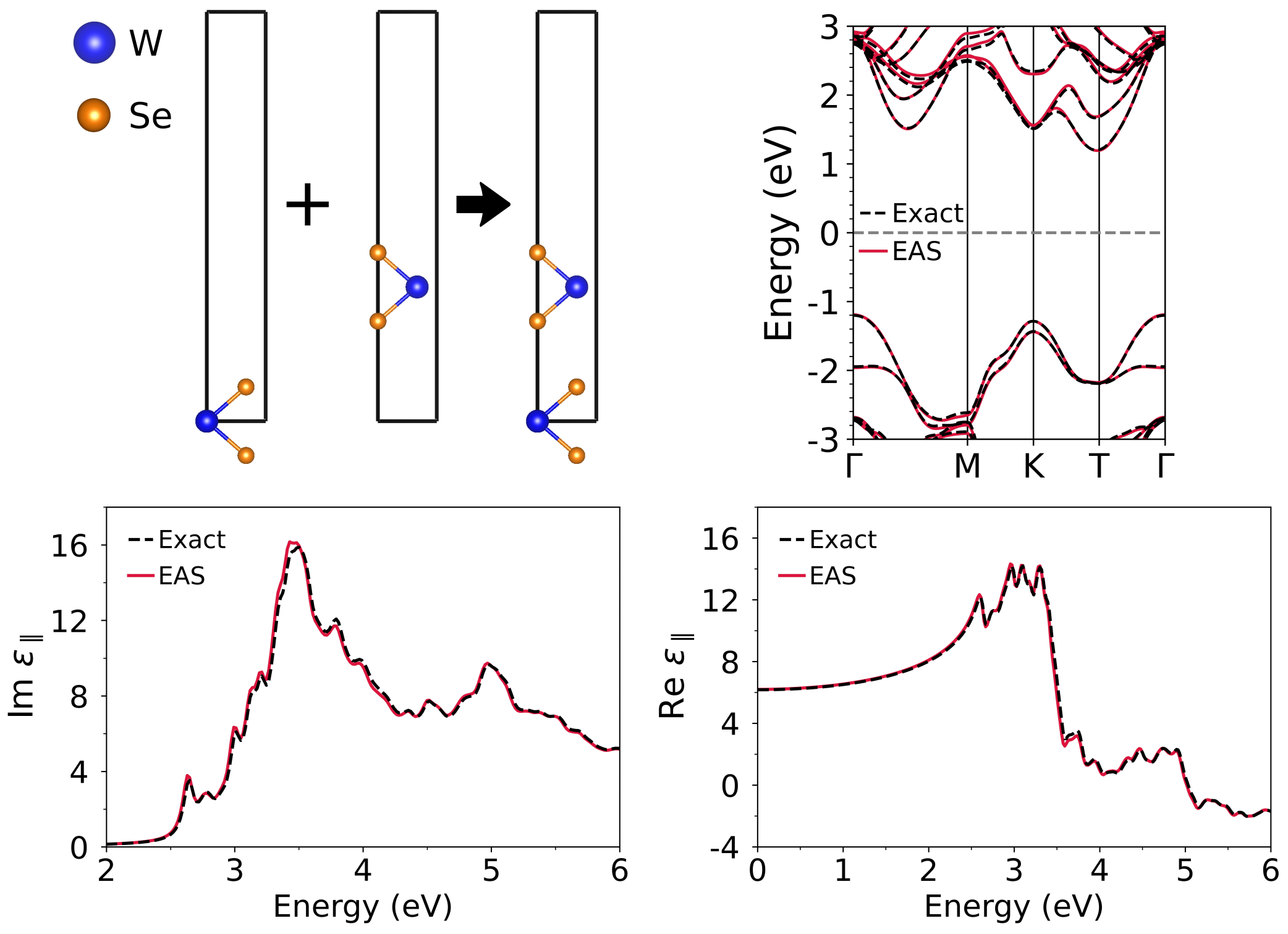}%
\caption{Top left: schematic representation of how to compute the additive screening, for the example of the WSe$_2$ bilayer. Top right: QP band structure calculated with the EAS method (red) compared to the exact calculation (black). Bottom: Imaginary part (left) and real part (right) of the in-plane component of the dielectric function obtained by the EAS approach (red) and the exact calculation (black).}
\label{fig:wse2_bilayer_eas}
\end{center}
\end{figure*}

\subsection{The EAS approach in the Bethe-Salpeter equation}
\label{section:eas_for_bse}

While the polarizability is a key component in $GW$ in the screening of the electron-electron interaction, it plays an analogous role in the screening of the electron-hole interaction in optical excitations. We therefore make use of the same concept and extend the EAS approach to the calculations of optical properties by means of the Bethe-Salpeter equation. Unlike the $GW$ implementation, the BSE implementation of \exciting\ currently employs a planewave representation of the polarizability. Therefore, the original EAS approach of Refs. \cite{Liu2019,Xuan2019} (Section~\ref{section:theoretical_background}) is directly applicable here. For more details on the BSE modules the reader is referred to Ref. \cite{Vorwerk2019}. 
    
\section{Results}
\label{section:results}

We have selected two systems to demonstrate our approach, which are the WSe$_2$ bilayer and the HIOS pyridine@MoS$_2$. Both are weakly bound systems, dominated by vdW interactions. To evaluate the accuracy of our approach, we compare the results of the exact calculations  for both \GW~and BSE calculations with the results obtained using EAS . We note that an additional example of the EAS method for the BSE has been shown in Ref.~\cite{GonzalezOliva2023} for pyrene@MoS$_2$. 

\subsection{WSe$_2$ bilayer}
\label{section:wse2_bilayer}

The top left panel of Fig.~\ref{fig:wse2_bilayer_eas} shows the schematic representation of how to compute the additive screening for the example of the WSe$_2$ bilayer. Since both layers  and thus the double layer contain one unit cell, we only probe the additive step in this case. We consider the AB stacking \citep{He2014} and the interlayer distance of 3.21 \AA, as obtained by optimizing the geometry using the PBE functional and the Tkatchenko-Scheffler (TS) vdW correction \cite{Tkatchenko2009,Tkatchenko2010}. The sampling of the Brillouin zone (BZ) is performed with a homogeneous 12$\times$12$\times$1 Monkhorst-Pack \textbf{k}-point grid, and we use 200 empty states for both the individual components and the heterostructure, which ensures that the band edges are overall sufficiently converged. We further employ a Coulomb truncation along the out-of-plane direction, as used in Ref. \cite{RodriguesPela2024}. Spin-orbit coupling is not considered.

\begin{figure*}[t!]
\begin{center}
\includegraphics[width=0.8\textwidth]{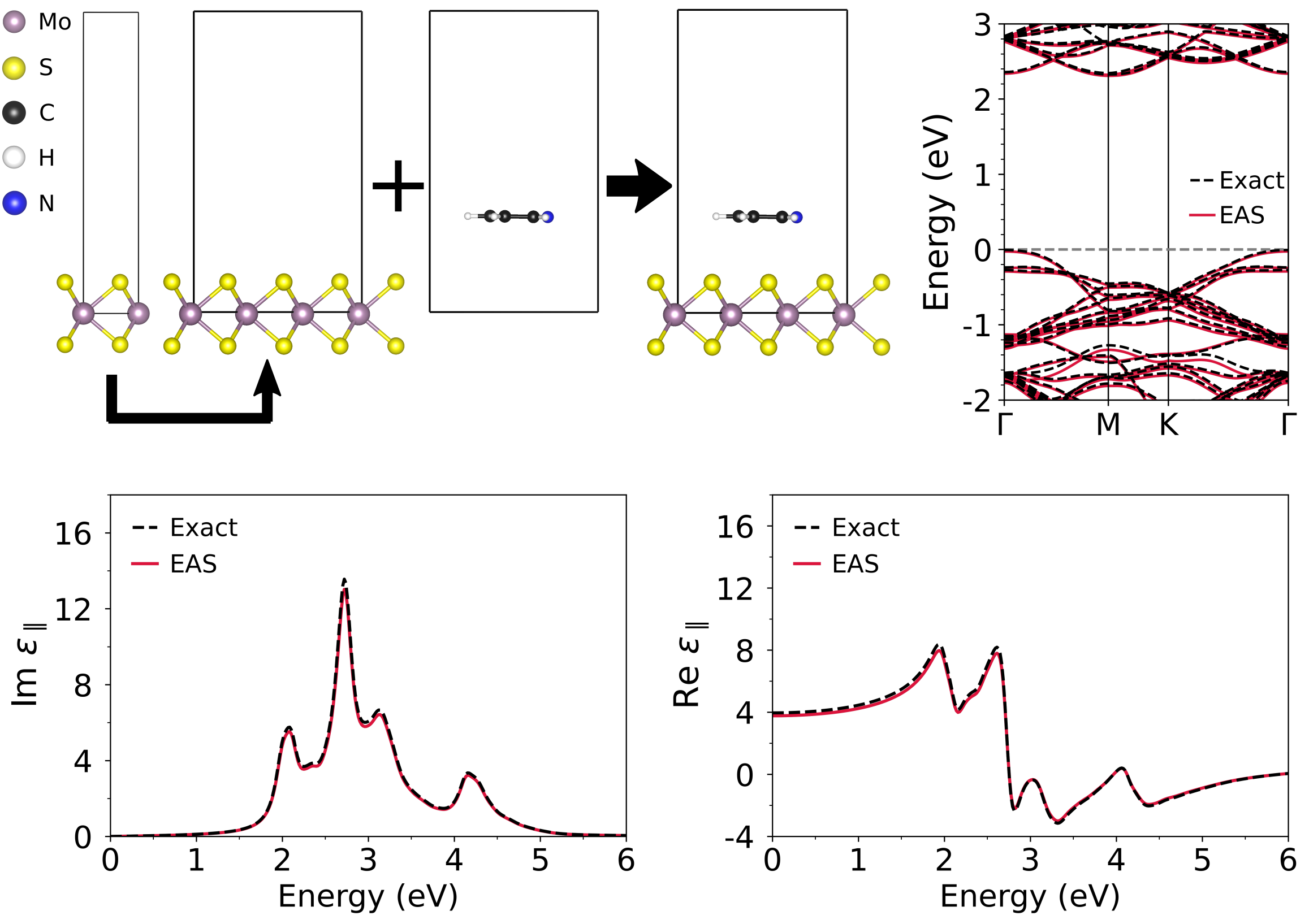}%
\caption{Top left: schematic representation of how to compute the EAS for the example of pyridine@MoS$_2$. Top right: QP band structure calculated with the EAS method (red) compared to the exact calculation (black). Bottom: Imaginary part (left) and real part (right) of the in-plane component of the dielectric function obtained by the EAS approach in both $GW$ and BSE (red) and the exact calculation (black).}
\label{fig:pyridine_mos2_eas}
\end{center}
\end{figure*}

The top right panel of Fig.~\ref{fig:wse2_bilayer_eas} shows the QP band structure of the bilayer computed by the EAS and compared to the the exact calculation. We observe a very good overall agreement in the entire energy range. The indirect fundamental gap, found between $\Gamma$ and T, has a value of 2.39 eV (2.41 eV) in the EAS (exact) calculation. The values of the direct gap at K are 2.84 eV and 2.79 eV, respectively. The bottom panels show the imaginary and real parts of the macroscopic dielectric function obtained with computing the polarizability in both $GW$ and BSE using the EAS approach. Again, we find very good agreement with the exact BSE calculation, finding optical gaps of 2.63 eV and 2.64 eV from the EAS and the exact calculation, respectively. 

We note that the obtained result show some discrepancies to other reported QP band gaps and optical gaps \cite{He2014}. We attribute this to the strong dependence of these calculations on structural parameters, such as the lattice constant, the interlayer distance, and the stacking, as well as on computational parameters, such as exchange-correlation functional, spin-orbit coupling, Coulomb truncation schemes, and the $\bq=0$ singularity treatment \cite{RodriguesPela2024,Haastrup_2018,Gjerding_2021}. However, the goal of this work is to show the applicability of the EAS method within one computational setup.

\subsection{Pyridine@MoS$_2$}
\label{section:pyridine_mos2}

The top left panel of Fig.~\ref{fig:pyridine_mos2_eas} shows the schematic representation of the EAS calculation for pyridine@MoS$_2$, a prototypical hybrid material. In this example, we are putting our full implementation to the test by performing an expansion of the polarizability calculated for the MoS$_2$ monolayer to a 3$\times$3$\times$1 supercell. We note that the expansion step is only exact if the supercell is an exact replica of the unit cell, \ie no geometry optimization of the heterostructure is performed. In the example shown here, the supercell is commensurate with the MoS$_2$ unit cell, with the atomic positions kept fixed at ideal lattice sites. The monolayer of pyridine is then placed at an interlayer distance of 3.19~\AA\ as obtained in Ref. \cite{GonzalezOliva2022} using the TS vdW correction. The BZ of the supercell is sampled with a 3$\times$3$\times$1 \textbf{k}-grid. As in the previous example, the calculation of the polarizability of the heterostructure is performed by using 200 empty bands, while the polarizability of the unit cell is computed using a 9$\times$9$\times$1 \textbf{k}-grid and 100 empty bands. Spin-orbit coupling is not considered.

\begin{table*}[t!]
\caption{\label{tab:computational_efficiency} Comparison of computing resources for pyridine@MoS$_2$ for both the exact method and the EAS screening method in the \GW/BSE framework. Steps that are shared between both methods are listed only once.}
\begin{ruledtabular}
\begin{tabular}{llcc}
\textbf{Framework} & \textbf{Task Name} & \textbf{CPU Hours (EAS)} & \textbf{CPU Hours (exact)} \\
\hline
\multirow{4}{*}{$G_0W_0$ }
& Initialization & 101 & 101 \\
& Polarizability & - & 3511 \\
& Polarizability MoS$_2$\footnotemark[2] & 260  & -  \\
& Polarizability pyridine unit cell & 1120& - \\
& Addition individual polarizabilities& 165 & -  \\
& Self-energy & 4055& 4055 \\
\cline{2-4}
& \textbf{Total} & 5701 & 7667 \\
\hline
\multirow{7}{*}{BSE}
& Initialization\footnotemark[1] & 557 & 557 \\
& Polarizability & - & 9462 \\
& Polarizability MoS$_2$\footnotemark[2] & 1122& - \\
& Polarizability pyridine unit cell & 1840 & - \\
& Addition individual polarizabilities& 0\footnotemark[3] & - \\
& Construction interaction Hamiltonian & 3824& 3824 \\
& Diagonalization Hamiltonian & 318& 318 \\
\cline{2-4}
& \textbf{Total} & 7661&  14161 \\
\end{tabular}
\end{ruledtabular}
\footnotetext[1]{Includes calculation of eigenvalues/eigenvectors and momentum matrix on the BSE \textbf{k}-grid.}
\footnotetext[2]{Includes the calculation of the unit cell polarizability and the expansion to the supercell.}
\footnotetext[3]{Rounded}
\end{table*}

In the top right panel of Fig.~\ref{fig:pyridine_mos2_eas} the QP band structure calculated with the EAS method (red) is compared with the exact calculation (black). As in the case of the WSe$_2$ bilayer, we find very good agreement. The fundamental gap appears at the $\Gamma$-point due to the folding of the BZ (K-point in the primitive cell). Its value of 2.32 eV comes very close to the result of the exact calculation of 2.36 eV. The optical properties, calculated on top of the QP band structures, indicate optical gaps of 2.00 eV (EAS) and 2.02 eV (exact), respectively. Figure~\ref{fig:pyridine_mos2_eas} depicts the imaginary and real parts of the macroscopic dielectric function of the corresponding spectra, for which we find an excellent agreement between the two approaches.

To demonstrate the computational efficiency of our approach, we provide the CPU hours~\footnote{CPU hours are defined as the number of processors used multiplied by the number of wall-time hours needed for the calculation.} required for the two methods in Table~\ref{tab:computational_efficiency}. We note that, as described in Ref. \cite{Liu2019}, the actual performance and timing depend on the architecture of the computer cluster used. Our calculations are performed using the computer clusters Lise and Emmy of the NHR@ZIB and NHR@Göttingen, both with Intel(R) Xeon(R) Silver 4210 CPU processors and 2.20 GHz frequency. The calculation of the exact polarizability in the  \GW~method for pyridine@MoS$_2$ requires 3511 CPU hours, while the calculations for the unit cells of MoS$_2$ and pyridine require 260 CPU hours and 1120 CPU hours, respectively. Including the addition step of 165 CPU hours brings the total runtime to 1545 CPU hours, representing a savings of over 50\% in computing resources.  In the BSE case, the exact method requires 9462 CPU hours for the calculation of the polarizability, while  the individual components require 1122 and 1840 CPU hours for MoS$_2$ and pyridine, respectively. This results in a total runtime of 2962 CPU hours, corresponding to a reduction of almost 70\% of the computational time. This even larger reduction in BSE as compared to the \GW case can be explained by the expensive basis transformations required for treating the mixed product basis.

We note that EAS only reduces the computational cost associated with the calculation of the polarizability. However, the calculation of the self-energy and the construction and diagonalization of the BSE Hamiltonian remain computationally demanding. Therefore, the total time is only reduced by 25\% and 46\% for \GW~and BSE, respectively. In the context of BSE, the combination of the EAS method with the interpolative separable density fitting (ISDF) method \cite{Henneke2020} represents an exceptionally compelling framework for the efficient calculation of optical properties of large weakly-bound hybrid systems. 

\section{Summary and Conclusions}
\label{section:conclusions}

In this work, we have extended a recently developed method for calculating the polarizability of weakly bound systems using a planewave basis \cite{Xuan2019,Liu2019} to the case of the (L)APW basis. This method consists of expressing the polarizability of the heterostructure as the sum of the polarizabilities of the individual components. Moreover, we have introduced a corresponding approach to compute the optical spectra of such heterostructures within the framework of the Bethe-Salpeter equation. We have developed all the necessary transformations in both the expansion and addition steps that are required to account for the atom-dependent nature of the mixed-product basis used in the context of the (L)APW method. While the expansion step requires the basis transformation between the unit cell and the supercell of the individual components, the addition step requires the transformation between the bases of the (expanded) individual components and the basis of the heterostructure.

Our implementation obtains the QP band structure of weakly bound heterostructures with high accuracy, as illustrated by the examples of the vdW heterostructure WSe$_2$ bilayer and the hybrid material pyridine@MoS$_2$. Going beyond previous work, we have successfully applied the EAS method also within the BSE formalism, which we have demonstrated by computing the optical spectra of the two systems. Due to its reduced computational cost, our method will be useful to provide highly accurate all-electron results of electronic and optical properties of heterostructures much larger than those considered so far. This opens the possibility to extend all-electron benchmarking efforts to more complex systems.

\section*{Acknowledgements}

This work was supported by Deutsche Forschungsgemeinschaft (DFG) within the Collaborative Research Center HIOS (SFB 951), project 182087777. M.S. acknowledges partial support from DFG Project 235221301. Partial support was also provided by the European Union’s Horizon 2020 Research and Innovation Program under the grant agreement no. 951786 (NOMAD CoE). I.G.O. thanks the DAAD (Deutscher Akademischer Austauschdienst) for financial support. Computing time on the supercomputers Lise and Emmy at NHR@ZIB and NHR@Göttingen is acknowledged. The authors are grateful to Sebastian Tillack and Andris Gulans for insightful discussions and Marti Raya-Moreno for critically reading the manuscript. \\

\appendix

\section{The (L)APW and mixed-product bases}
\label{appendix:mixed_basis}

In the (L)APW method, the unit cell is partitioned into the {\rm MT} and I regions and the basis is defined as
 \begin{widetext}
\begin{numcases}{\phi^{\bk}_{\bG} (\br) =}
    \sum_{\zeta lm} \mathbf{A}_{\alpha \zeta lm } (\bk + \bG) \, u_{\alpha \zeta l} (r^{\alpha}, \epsilon_{\alpha l}) \, Y_{lm} (\hat{\mathbf{r}}^{\alpha}) & $r^{\alpha} < R^{\alpha}_{\text{{\rm MT}}}$ \:, \label{eq:lapw_mt} \\
    \frac{1}{\sqrt{\Omega}} \, e^{\text{i}(\bk + \bG) \cdot \br} & $\br \in \text{I}$ \:. \label{eq:lapw_int}
\end{numcases}
\end{widetext}
The {\rm MT} part of $\phi^{\bk}_{\bG} (\br)$ is expanded in terms of spherical harmonics $Y_{lm} (\hat{\mathbf{r}}^{\alpha})$ and radial functions $u_{\alpha \zeta l} (r^{\alpha}, \epsilon_{\alpha l})$. The latter are solutions of the radial scalar-relativistic Schrödinger equation at ﬁxed reference energies $l\alpha$ (linearization energies) inside each {\rm MT} for the respective spherically averaged Kohn-Sham potential. The index $\zeta$ has been introduced to account for different types of radial functions, which are either $u_{\alpha l} (r^{\alpha}, \epsilon_{\alpha l})$ or $\dot{u}_{\alpha l} (r^{\alpha}, \epsilon_{\alpha l}) \equiv \frac{\partial u_{\alpha l} (r^{\alpha}, E)  }{\partial E} \left. \right|_{\epsilon_{\alpha l}}$ in the LAPW method. The augmentation coefﬁcients $\mathbf{A}_{\alpha \zeta lm } (\bk + \bG) $ are chosen such to ensure that the basis functions are continuous at the boundaries of the MT spheres. 

To represent products of two Kohn–Sham wavefunctions, for example, in the \GW~method, a basis with similar features as the (L)APW basis should be used \cite{Jiang2013}. In particular, one can make use of the fact that in the MT spheres, the product of two spherical harmonics can again be expanded in spherical harmonics using Clebsch-Gordan coefficients, while in the interstitial region the product of two planewaves is a planewave. Therefore, within the {\rm MT} region, an optimal set of radial functions $v_{\alpha NL} (r_{\alpha})$ is constructed from the products of the radial wavefunctions $u_{\alpha l} (r^{\alpha}, \epsilon_{\alpha l})$. We calculate the overlap matrix between these products functions,
\begin{widetext}
\begin{align}\label{eq:overlap_radial_product}
    \mathbb{O}_{ll';l_1 l_1'} = \int_0^{R_{{\rm MT}}^{\alpha}} \text{d}r^{\alpha} \, u_{\alpha l} (r^{\alpha}) \, u_{\alpha l'} (r^{\alpha}) \, u_{\alpha l_1} (r^{\alpha}) \, u_{\alpha l_1'} (r^{\alpha})(r^{\alpha})^2  \:,
\end{align}
\end{widetext}

which is then diagonalized to give the corresponding set of eigenvalues, $\lambda_N^{MB}$, and eigenvectors, $c_{ll',N}$. The eigenvectors are normalized and form the radial basis set according to 
\begin{align}\label{eq:radial_basis_set}
    v_{\alpha NL} (r^{\alpha}) = \sum_{ll'} c_{ll',N} \, u_{l} (r^{\alpha}) \, u_{l'} (r^{\alpha}) \:.
\end{align}

In the interstitial region, the overlap of two planewaves is given by
\begin{align}\label{eq:overlap_planewave_product}
    \mathbb{O}_{\bG \bG'} = \frac{1}{\Omega} \int_{\Omega} \text{d} \textbf{r} \, \theta_I (\br) \, e^{\text{i}(\bG - \bG')\cdot \br}  \:,
\end{align}
where $\theta_{I} (\br)$ is the Heaviside function which has a value of 1 in the interstitial region and zero otherwise. To obtain a set of orthonormal basis function, the overlap matrix is then diagonalized and the basis is defined in terms of the eigenvalues $\lambda_{i}^{I}$ and eigenvectors $S_{\bG i}$ according to Eq.~\ref{eq:int_mpb} using the normalized eigenvectors defined as
\begin{align}\label{eq:planewave_eigenvector}
    \Tilde{S}_{\bG i} \equiv \frac{S_{\bG i}}{\sqrt{\lambda_i^I}} \:.
\end{align}

Finally, we note that to construct the transformation shown in the addition steps, the \exciting\ code takes advantage of the planewave representation using the overlap integrals defined as
\begin{align}\label{eq:overlap_integrals}
    w_{i\bG} (\bq) \equiv \frac{1}{\Omega} \int_{\Omega} \text{d}\br \, \Big[ \chi^{\bq}_i (\br) \Big]^* \, e^{\text{i}(\bq+\bG) \cdot \br}  \:.
\end{align}

Since we have already defined $\chi^{\bq}_i (\br)$ in both the MT region, $\chi^{\bq}_i (\br) \rightarrow \gamma^{\bq}_{\alpha NLM} (\br)$, and the interstitial region, $\chi^{\bq}_i (\br) \rightarrow \Lambda^{q}_{i} (\br)$, Eq.~\ref{eq:overlap_integrals} allows us to transform between the two. For a detailed overview of the mixed-product basis based on the (L)APW method, the reader is referred to Refs.~\cite{Jiang2013} and~\cite{Nabok2016}.

\section{Calculation of the transformation matrices}
\label{appendix:overlap_matrix}

As part of the addition step (Section~\ref{section:addition_step}) we define the matrices $C^\mu_{aA}(\bq)=\braket{\chi^{\mu,\bq}_a}{\chi_A^{\bq}}$ to transform the coefficients of the polarizability of a component to the basis of the full heterostructure. Using Eq.~\ref{eq:mat_overlap}, we construct the overlap matrix for the case of two-layered system,
\begin{align}\label{eq:mat_overlap_c1}
    C^1 = &
\begin{pmatrix}
    C_{{\rm MT}_1/{\rm MT}_1} & C_{{\rm MT}_1/{\rm MT}_2} & C_{{\rm MT}_1/I} \\[5pt]
    C_{I_1/{\rm MT}_1} & C_{I_1/{\rm MT}_2} & C_{I_1/I}
\end{pmatrix} \:, \nonumber \\[5pt]
= &
\begin{pmatrix}
    1 & 0 & 0 \\[5pt]
    0 & C_{I_1/{\rm MT}_2} & C_{I_1/I}
\end{pmatrix} \:,
\end{align}
and
\begin{align}\label{eq:mat_overlap_c2}
    C^2 = &
\begin{pmatrix}
    C_{{\rm MT}_2/{\rm MT}_1} & C_{{\rm MT}_2/{\rm MT}_2} & C_{{\rm MT}_2/I} \\[5pt]
    C_{I_2/{\rm MT}_1} & C_{I_2/{\rm MT}_2} & C_{I_2/I}
\end{pmatrix} \:, \nonumber \\[5pt]
= &
\begin{pmatrix}
    0 & 1 & 0 \\[5pt]
    C_{I_2/{\rm MT}_1} & 0 & C_{I_2/I}
\end{pmatrix} \:,
\end{align}
where the zero terms have been obtained by using the conditions of Section~\ref{section:addition_step}. In component 1, Eq.~\ref{eq:mat_overlap_c1}, the only non-trivial contributions to compute are $C_{I_1/{\rm MT}_2}$ and $C_{I_1/I}$. In the following, we show how to compute them in terms of the planewave eigenvectors (Eq.~\ref{eq:planewave_eigenvector}) and the overlap matrices (Eq.~\ref{eq:overlap_integrals}). We begin with $C_{I_1/{\rm MT}_2}$, where one element of this matrix is given by
\begin{widetext}
\begin{align}\label{eq:C_MT/I}
    C_{i\alpha} & =  \frac{1}{\sqrt{\Omega}} \, \int_{{\rm MT}} \text{d}\br
                \left[\sum_{\bG}\tilde{S}^{*}_{\bG i} \, e^{\text{i}(\bq+\bG)\br}\right]^{\! *}
                \left[\sum_\bR e^{-\text{i} \, \bq(\bR+\br_\alpha)} \,
                    v_{\alpha NL}(r^\alpha) \, Y^*_{LM}(\hat{\br}^\alpha) \right]^* \:, \nonumber 
               \\[5pt]
    &= \sum_{\bG}\tilde{S}_{\bG i} \, w^{*}_{\alpha \bG}(\bq) \:,
\end{align}
\end{widetext}
where $w_{\alpha \bG}(\bq)$ is the overlap element of a planewave with the basis functions in the {\rm MT} region of the heterostructure. 

The other non-zero component of Eq.~\ref{eq:mat_overlap_c1}, $C_{I_1/I}$ is constructed by considering the basis representation of the interstitial region of the individual component and the heterostructure. Therefore one element of this matrix is given by
\begin{align}
    C_{iI} & = \frac{1}{\Omega}\int_{I_{HS}} \text{d}\br\sum_{\bG'\bG}
    \tilde{S}_{\bG' i} \, e^{-\text{i}(\bq+\bG')\br} \,
    \tilde{S}^*_{\bG I} \, e^{\text{i}(\bq+\bG)\br} \:, \nonumber \\[5pt]
    & = \frac{1}{\Omega} \sum_{\bG'\bG}
    \tilde{S}_{\bG' i} \, \tilde{S}^*_{\bG I}\int_{I_{HS}} \text{d}\br \, e^{\text{i}(\bG-\bG')\br} \:, \nonumber \\[5pt]
    & = \sum_{\bG'}\tilde{S}_{\bG' i} \, \frac{1}{\Omega}\sum_{\bG}\tilde{S}^*_{\bG I}\int_{I_{HS}} \text{d}\br \, e^{\text{i}(\bG-\bG')\br} .
\end{align}

We define $\mathcal{I}_\bG$ as the integral over a planewave evaluated for the interstitial region of the heterostructure as
\begin{equation}
    \mathcal{I}_\bG = \int_{I_{\Omega}} \, \text{d}\br \, e^{\text{i} \bG\br} \:,
\end{equation}
such that we can express the matrix elements as
\begin{align}\label{eq:C_I/I}
    C_{iI} & =   \sum_{\bG'}\tilde{S}_{\bG' i} \, \frac{1}{\Omega} \sum_{\bG}
    \tilde{S}^*_{\bG I} \, \mathcal{I}_{\bG - \bG'} \:, \nonumber \\[5pt]
    & = \sum_{\bG'}\tilde{S}_{\bG' i} \, \left[ \frac{1}{\Omega}\sum_{\bG}
    \tilde{S}_{\bG I} \, \mathcal{I}_{\bG' - \bG} \right]^* \:, \nonumber  \\[5pt]
    & =  \sum_{\bG'}\tilde{S}_{\bG' i} \left[ \frac{1}{\Omega}\sum_{\bG}
    \tilde{S}_{\bG I} \, \mathcal{I}^*_{\bG - \bG'} \right]^* \:, \nonumber \\[5pt]
     & = \sum_{\bG'}\tilde{S}_{\bG' i} \, w^*_{I\bG'} \:.
\end{align}
where $w_{I\bG}(\bq)$ is the overlap element of a planewave with the basis functions in the interstitial region of the heterostructure.

We conclude this section by noting that the transformation of component 2 can be achieved using Eq.~\ref{eq:C_MT/I} for $C_{I_2/{\rm MT}_1}$ and Eq.~\ref{eq:C_I/I} for $C_{I_2/I}$.

\bibliography{article.bib}

\begin{thebibliography}{42}%
\makeatletter
\providecommand \@ifxundefined [1]{%
 \@ifx{#1\undefined}
}%
\providecommand \@ifnum [1]{%
 \ifnum #1\expandafter \@firstoftwo
 \else \expandafter \@secondoftwo
 \fi
}%
\providecommand \@ifx [1]{%
 \ifx #1\expandafter \@firstoftwo
 \else \expandafter \@secondoftwo
 \fi
}%
\providecommand \natexlab [1]{#1}%
\providecommand \enquote  [1]{``#1''}%
\providecommand \bibnamefont  [1]{#1}%
\providecommand \bibfnamefont [1]{#1}%
\providecommand \citenamefont [1]{#1}%
\providecommand \href@noop [0]{\@secondoftwo}%
\providecommand \href [0]{\begingroup \@sanitize@url \@href}%
\providecommand \@href[1]{\@@startlink{#1}\@@href}%
\providecommand \@@href[1]{\endgroup#1\@@endlink}%
\providecommand \@sanitize@url [0]{\catcode `\\12\catcode `\$12\catcode `\&12\catcode `\#12\catcode `\^12\catcode `\_12\catcode `\%12\relax}%
\providecommand \@@startlink[1]{}%
\providecommand \@@endlink[0]{}%
\providecommand \url  [0]{\begingroup\@sanitize@url \@url }%
\providecommand \@url [1]{\endgroup\@href {#1}{\urlprefix }}%
\providecommand \urlprefix  [0]{URL }%
\providecommand \Eprint [0]{\href }%
\providecommand \doibase [0]{http://dx.doi.org/}%
\providecommand \selectlanguage [0]{\@gobble}%
\providecommand \bibinfo  [0]{\@secondoftwo}%
\providecommand \bibfield  [0]{\@secondoftwo}%
\providecommand \translation [1]{[#1]}%
\providecommand \BibitemOpen [0]{}%
\providecommand \bibitemStop [0]{}%
\providecommand \bibitemNoStop [0]{.\EOS\space}%
\providecommand \EOS [0]{\spacefactor3000\relax}%
\providecommand \BibitemShut  [1]{\csname bibitem#1\endcsname}%
\let\auto@bib@innerbib\@empty
\bibitem [{\citenamefont {Wang}\ \emph {et~al.}(2021)\citenamefont {Wang}, \citenamefont {Jia}, \citenamefont {Huang},\ and\ \citenamefont {Duan}}]{Wang2021_review}%
  \BibitemOpen
  \bibfield  {author} {\bibinfo {author} {\bibfnamefont {P.}~\bibnamefont {Wang}}, \bibinfo {author} {\bibfnamefont {C.}~\bibnamefont {Jia}}, \bibinfo {author} {\bibfnamefont {Y.}~\bibnamefont {Huang}}, \ and\ \bibinfo {author} {\bibfnamefont {X.}~\bibnamefont {Duan}},\ }\href {\doibase https://doi.org/10.1016/j.matt.2020.12.015} {\bibfield  {journal} {\bibinfo  {journal} {Matter}\ }\textbf {\bibinfo {volume} {4}},\ \bibinfo {pages} {552} (\bibinfo {year} {2021})}\BibitemShut {NoStop}%
\bibitem [{\citenamefont {Molina-S{\'{a}}nchez}\ \emph {et~al.}(2015)\citenamefont {Molina-S{\'{a}}nchez}, \citenamefont {Hummer},\ and\ \citenamefont {Wirtz}}]{Molina2015}%
  \BibitemOpen
  \bibfield  {author} {\bibinfo {author} {\bibfnamefont {A.}~\bibnamefont {Molina-S{\'{a}}nchez}}, \bibinfo {author} {\bibfnamefont {K.}~\bibnamefont {Hummer}}, \ and\ \bibinfo {author} {\bibfnamefont {L.}~\bibnamefont {Wirtz}},\ }\href {\doibase 10.1016/j.surfrep.2015.10.001} {\bibfield  {journal} {\bibinfo  {journal} {Surf.~Sci.~Rep.~}\ }\textbf {\bibinfo {volume} {70}},\ \bibinfo {pages} {554} (\bibinfo {year} {2015})}\BibitemShut {NoStop}%
\bibitem [{\citenamefont {Wang}\ \emph {et~al.}(2018)\citenamefont {Wang}, \citenamefont {Chernikov}, \citenamefont {Glazov}, \citenamefont {Heinz}, \citenamefont {Marie}, \citenamefont {Amand},\ and\ \citenamefont {Urbaszek}}]{Wang2018}%
  \BibitemOpen
  \bibfield  {author} {\bibinfo {author} {\bibfnamefont {G.}~\bibnamefont {Wang}}, \bibinfo {author} {\bibfnamefont {A.}~\bibnamefont {Chernikov}}, \bibinfo {author} {\bibfnamefont {M.~M.}\ \bibnamefont {Glazov}}, \bibinfo {author} {\bibfnamefont {T.~F.}\ \bibnamefont {Heinz}}, \bibinfo {author} {\bibfnamefont {X.}~\bibnamefont {Marie}}, \bibinfo {author} {\bibfnamefont {T.}~\bibnamefont {Amand}}, \ and\ \bibinfo {author} {\bibfnamefont {B.}~\bibnamefont {Urbaszek}},\ }\href {\doibase 10.1103/RevModPhys.90.021001} {\bibfield  {journal} {\bibinfo  {journal} {Rev.~Mod.~Phys.~}\ }\textbf {\bibinfo {volume} {90}},\ \bibinfo {pages} {021001} (\bibinfo {year} {2018})}\BibitemShut {NoStop}%
\bibitem [{\citenamefont {Devakul}\ \emph {et~al.}(2021)\citenamefont {Devakul}, \citenamefont {Cr{\'e}pel}, \citenamefont {Zhang},\ and\ \citenamefont {Fu}}]{Devakul2021-gi}%
  \BibitemOpen
  \bibfield  {author} {\bibinfo {author} {\bibfnamefont {T.}~\bibnamefont {Devakul}}, \bibinfo {author} {\bibfnamefont {V.}~\bibnamefont {Cr{\'e}pel}}, \bibinfo {author} {\bibfnamefont {Y.}~\bibnamefont {Zhang}}, \ and\ \bibinfo {author} {\bibfnamefont {L.}~\bibnamefont {Fu}},\ }\href {https://doi.org/10.1038/s41467-021-27042-9} {\bibfield  {journal} {\bibinfo  {journal} {Nat. Commun.}\ }\textbf {\bibinfo {volume} {12}},\ \bibinfo {pages} {6730} (\bibinfo {year} {2021})}\BibitemShut {NoStop}%
\bibitem [{\citenamefont {Caruso}\ \emph {et~al.}(2022)\citenamefont {Caruso}, \citenamefont {Schebek}, \citenamefont {Pan}, \citenamefont {Vona},\ and\ \citenamefont {Draxl}}]{Caruso2022}%
  \BibitemOpen
  \bibfield  {author} {\bibinfo {author} {\bibfnamefont {F.}~\bibnamefont {Caruso}}, \bibinfo {author} {\bibfnamefont {M.}~\bibnamefont {Schebek}}, \bibinfo {author} {\bibfnamefont {Y.}~\bibnamefont {Pan}}, \bibinfo {author} {\bibfnamefont {C.}~\bibnamefont {Vona}}, \ and\ \bibinfo {author} {\bibfnamefont {C.}~\bibnamefont {Draxl}},\ }\href {\doibase 10.1021/acs.jpclett.2c01034} {\bibfield  {journal} {\bibinfo  {journal} {The Journal of Physical Chemistry Letters}\ }\textbf {\bibinfo {volume} {13}},\ \bibinfo {pages} {5894–5899} (\bibinfo {year} {2022})}\BibitemShut {NoStop}%
\bibitem [{\citenamefont {Ji}\ and\ \citenamefont {Choi}(2022)}]{Ji2022}%
  \BibitemOpen
  \bibfield  {author} {\bibinfo {author} {\bibfnamefont {J.}~\bibnamefont {Ji}}\ and\ \bibinfo {author} {\bibfnamefont {J.~H.}\ \bibnamefont {Choi}},\ }\href {\doibase http://dx.doi.org/10.1039/D2NR01358D} {\bibfield  {journal} {\bibinfo  {journal} {Nanoscale}\ }\textbf {\bibinfo {volume} {14}},\ \bibinfo {pages} {10648} (\bibinfo {year} {2022})}\BibitemShut {NoStop}%
\bibitem [{\citenamefont {Draxl}\ \emph {et~al.}(2014)\citenamefont {Draxl}, \citenamefont {Nabok},\ and\ \citenamefont {Hannewald}}]{Draxl2014}%
  \BibitemOpen
  \bibfield  {author} {\bibinfo {author} {\bibfnamefont {C.}~\bibnamefont {Draxl}}, \bibinfo {author} {\bibfnamefont {D.}~\bibnamefont {Nabok}}, \ and\ \bibinfo {author} {\bibfnamefont {K.}~\bibnamefont {Hannewald}},\ }\href {\doibase 10.1021/ar500096q} {\bibfield  {journal} {\bibinfo  {journal} {Acc.~Chem.~Res.~}\ }\textbf {\bibinfo {volume} {47}},\ \bibinfo {pages} {3225} (\bibinfo {year} {2014})}\BibitemShut {NoStop}%
\bibitem [{\citenamefont {Hofmann}\ \emph {et~al.}(2021)\citenamefont {Hofmann}, \citenamefont {Zojer}, \citenamefont {Hörmann}, \citenamefont {Jeindl},\ and\ \citenamefont {Maurer}}]{Hofmann2021}%
  \BibitemOpen
  \bibfield  {author} {\bibinfo {author} {\bibfnamefont {O.~T.}\ \bibnamefont {Hofmann}}, \bibinfo {author} {\bibfnamefont {E.}~\bibnamefont {Zojer}}, \bibinfo {author} {\bibfnamefont {L.}~\bibnamefont {Hörmann}}, \bibinfo {author} {\bibfnamefont {A.}~\bibnamefont {Jeindl}}, \ and\ \bibinfo {author} {\bibfnamefont {R.~J.}\ \bibnamefont {Maurer}},\ }\href {\doibase 10.1039/D0CP06605B} {\bibfield  {journal} {\bibinfo  {journal} {Phys. Chem. Chem. Phys.}\ }\textbf {\bibinfo {volume} {23}},\ \bibinfo {pages} {8132} (\bibinfo {year} {2021})}\BibitemShut {NoStop}%
\bibitem [{\citenamefont {Gonzalez~Oliva}\ \emph {et~al.}(2024)\citenamefont {Gonzalez~Oliva}, \citenamefont {Maurer}, \citenamefont {Alex}, \citenamefont {Tillack}, \citenamefont {Schebek},\ and\ \citenamefont {Draxl}}]{GonzalezOliva2023}%
  \BibitemOpen
  \bibfield  {author} {\bibinfo {author} {\bibfnamefont {I.}~\bibnamefont {Gonzalez~Oliva}}, \bibinfo {author} {\bibfnamefont {B.}~\bibnamefont {Maurer}}, \bibinfo {author} {\bibfnamefont {B.}~\bibnamefont {Alex}}, \bibinfo {author} {\bibfnamefont {S.}~\bibnamefont {Tillack}}, \bibinfo {author} {\bibfnamefont {M.}~\bibnamefont {Schebek}}, \ and\ \bibinfo {author} {\bibfnamefont {C.}~\bibnamefont {Draxl}},\ }\href {\doibase 10.1002/pssa.202300170} {\bibfield  {journal} {\bibinfo  {journal} {Phys.~Status~Solidi~A}\ }\textbf {\bibinfo {volume} {221}},\ \bibinfo {pages} {2300170} (\bibinfo {year} {2024})}\BibitemShut {NoStop}%
\bibitem [{\citenamefont {Liu}(2025)}]{Liu2025}%
  \BibitemOpen
  \bibfield  {author} {\bibinfo {author} {\bibfnamefont {Z.-F.}\ \bibnamefont {Liu}},\ }\href {\doibase 10.1021/acsnano.4c18268} {\bibfield  {journal} {\bibinfo  {journal} {ACS Nano}\ }\textbf {\bibinfo {volume} {19}},\ \bibinfo {pages} {5861} (\bibinfo {year} {2025})}\BibitemShut {NoStop}%
\bibitem [{\citenamefont {Benson}\ and\ \citenamefont {Koch}(2024)}]{Benson2024}%
  \BibitemOpen
  \bibfield  {author} {\bibinfo {author} {\bibfnamefont {O.}~\bibnamefont {Benson}}\ and\ \bibinfo {author} {\bibfnamefont {N.}~\bibnamefont {Koch}},\ }\href {\doibase https://doi.org/10.1002/pssa.202300939} {\bibfield  {journal} {\bibinfo  {journal} {Phys. Status Solidi A}\ }\textbf {\bibinfo {volume} {221}},\ \bibinfo {pages} {2300939} (\bibinfo {year} {2024})}\BibitemShut {NoStop}%
\bibitem [{\citenamefont {Xuan}\ \emph {et~al.}(2019)\citenamefont {Xuan}, \citenamefont {Chen},\ and\ \citenamefont {Quek}}]{Xuan2019}%
  \BibitemOpen
  \bibfield  {author} {\bibinfo {author} {\bibfnamefont {F.}~\bibnamefont {Xuan}}, \bibinfo {author} {\bibfnamefont {Y.}~\bibnamefont {Chen}}, \ and\ \bibinfo {author} {\bibfnamefont {S.~Y.}\ \bibnamefont {Quek}},\ }\href {\doibase 10.1021/acs.jctc.9b00229} {\bibfield  {journal} {\bibinfo  {journal} {Journal of Chemical Theory and Computation}\ }\textbf {\bibinfo {volume} {15}},\ \bibinfo {pages} {3824} (\bibinfo {year} {2019})}\BibitemShut {NoStop}%
\bibitem [{\citenamefont {Liu}\ \emph {et~al.}(2019)\citenamefont {Liu}, \citenamefont {Jornada}, \citenamefont {Louie},\ and\ \citenamefont {Neaton}}]{Liu2019}%
  \BibitemOpen
  \bibfield  {author} {\bibinfo {author} {\bibfnamefont {Z.~F.}\ \bibnamefont {Liu}}, \bibinfo {author} {\bibfnamefont {F.~H.~D.}\ \bibnamefont {Jornada}}, \bibinfo {author} {\bibfnamefont {S.~G.}\ \bibnamefont {Louie}}, \ and\ \bibinfo {author} {\bibfnamefont {J.~B.}\ \bibnamefont {Neaton}},\ }\href {\doibase 10.1021/acs.jctc.9b00326} {\bibfield  {journal} {\bibinfo  {journal} {Journal of Chemical Theory and Computation}\ }\textbf {\bibinfo {volume} {15}},\ \bibinfo {pages} {4218} (\bibinfo {year} {2019})}\BibitemShut {NoStop}%
\bibitem [{\citenamefont {Adeniran}\ and\ \citenamefont {Liu}(2021)}]{Adeniran2021}%
  \BibitemOpen
  \bibfield  {author} {\bibinfo {author} {\bibfnamefont {O.}~\bibnamefont {Adeniran}}\ and\ \bibinfo {author} {\bibfnamefont {Z.~F.}\ \bibnamefont {Liu}},\ }\href {\doibase 10.1063/5.0072995} {\bibfield  {journal} {\bibinfo  {journal} {J.~Chem.~Phys.~}\ }\textbf {\bibinfo {volume} {155}},\ \bibinfo {pages} {214702} (\bibinfo {year} {2021})}\BibitemShut {NoStop}%
\bibitem [{\citenamefont {Frimpong}\ and\ \citenamefont {Liu}(2024)}]{Frimpong2024}%
  \BibitemOpen
  \bibfield  {author} {\bibinfo {author} {\bibfnamefont {J.}~\bibnamefont {Frimpong}}\ and\ \bibinfo {author} {\bibfnamefont {Z.~F.}\ \bibnamefont {Liu}},\ }\href {\doibase 10.1021/acs.jpclett.3c03470} {\bibfield  {journal} {\bibinfo  {journal} {J.~Phys.~Chem.~Lett.}\ }\textbf {\bibinfo {volume} {15}},\ \bibinfo {pages} {2133} (\bibinfo {year} {2024})}\BibitemShut {NoStop}%
\bibitem [{\citenamefont {Palummo}\ \emph {et~al.}(2019)\citenamefont {Palummo}, \citenamefont {D'Auria}, \citenamefont {Grossman},\ and\ \citenamefont {Cicero}}]{Palummo2019}%
  \BibitemOpen
  \bibfield  {author} {\bibinfo {author} {\bibfnamefont {M.}~\bibnamefont {Palummo}}, \bibinfo {author} {\bibfnamefont {A.~N.}\ \bibnamefont {D'Auria}}, \bibinfo {author} {\bibfnamefont {J.~C.}\ \bibnamefont {Grossman}}, \ and\ \bibinfo {author} {\bibfnamefont {G.}~\bibnamefont {Cicero}},\ }\href {\doibase 10.1088/1361-648X/ab0c5e} {\bibfield  {journal} {\bibinfo  {journal} {J.~Phys.~Condens.~Matter.~}\ }\textbf {\bibinfo {volume} {31}},\ \bibinfo {pages} {235701} (\bibinfo {year} {2019})}\BibitemShut {NoStop}%
\bibitem [{\citenamefont {Wang}\ and\ \citenamefont {Paulus}(2020)}]{Wang2020}%
  \BibitemOpen
  \bibfield  {author} {\bibinfo {author} {\bibfnamefont {K.}~\bibnamefont {Wang}}\ and\ \bibinfo {author} {\bibfnamefont {B.}~\bibnamefont {Paulus}},\ }\href {\doibase 10.1039/d0cp01239d} {\bibfield  {journal} {\bibinfo  {journal} {Phys.~Chem.~Chem.~Phys.~}\ }\textbf {\bibinfo {volume} {22}},\ \bibinfo {pages} {11936} (\bibinfo {year} {2020})}\BibitemShut {NoStop}%
\bibitem [{\citenamefont {Wang}\ and\ \citenamefont {Paulus}(2021)}]{Wang2021}%
  \BibitemOpen
  \bibfield  {author} {\bibinfo {author} {\bibfnamefont {K.}~\bibnamefont {Wang}}\ and\ \bibinfo {author} {\bibfnamefont {B.}~\bibnamefont {Paulus}},\ }\href {\doibase 10.1021/acs.jpcc.1c05473} {\bibfield  {journal} {\bibinfo  {journal} {J.~Phys.~Chem.~C}\ }\textbf {\bibinfo {volume} {125}},\ \bibinfo {pages} {19544} (\bibinfo {year} {2021})}\BibitemShut {NoStop}%
\bibitem [{\citenamefont {Gonzalez~Oliva}\ \emph {et~al.}(2022)\citenamefont {Gonzalez~Oliva}, \citenamefont {Caruso}, \citenamefont {Pavone},\ and\ \citenamefont {Draxl}}]{GonzalezOliva2022}%
  \BibitemOpen
  \bibfield  {author} {\bibinfo {author} {\bibfnamefont {I.}~\bibnamefont {Gonzalez~Oliva}}, \bibinfo {author} {\bibfnamefont {F.}~\bibnamefont {Caruso}}, \bibinfo {author} {\bibfnamefont {P.}~\bibnamefont {Pavone}}, \ and\ \bibinfo {author} {\bibfnamefont {C.}~\bibnamefont {Draxl}},\ }\href {\doibase 10.1103/PhysRevMaterials.6.054004} {\bibfield  {journal} {\bibinfo  {journal} {Phys.~Rev.~Mater.~}\ }\textbf {\bibinfo {volume} {6}},\ \bibinfo {pages} {054004} (\bibinfo {year} {2022})}\BibitemShut {NoStop}%
\bibitem [{\citenamefont {Hybertsen}\ and\ \citenamefont {Louie}(1985)}]{Hybertsen1985}%
  \BibitemOpen
  \bibfield  {author} {\bibinfo {author} {\bibfnamefont {M.~S.}\ \bibnamefont {Hybertsen}}\ and\ \bibinfo {author} {\bibfnamefont {S.~G.}\ \bibnamefont {Louie}},\ }\href {\doibase 10.1103/PhysRevLett.55.1418} {\bibfield  {journal} {\bibinfo  {journal} {Phys.~Rev.~Lett.~}\ }\textbf {\bibinfo {volume} {55}},\ \bibinfo {pages} {1418} (\bibinfo {year} {1985})}\BibitemShut {NoStop}%
\bibitem [{\citenamefont {Hybertsen}\ and\ \citenamefont {Louie}(1986)}]{Hybertsen1986}%
  \BibitemOpen
  \bibfield  {author} {\bibinfo {author} {\bibfnamefont {M.~S.}\ \bibnamefont {Hybertsen}}\ and\ \bibinfo {author} {\bibfnamefont {S.~G.}\ \bibnamefont {Louie}},\ }\href {\doibase 10.1103/PhysRevB.34.5390} {\bibfield  {journal} {\bibinfo  {journal} {Phys.~Rev.~B}\ }\textbf {\bibinfo {volume} {34}},\ \bibinfo {pages} {5390} (\bibinfo {year} {1986})}\BibitemShut {NoStop}%
\bibitem [{\citenamefont {Liu}\ \emph {et~al.}(2016)\citenamefont {Liu}, \citenamefont {Kaltak}, \citenamefont {Klime\ifmmode~\check{s}\else \v{s}\fi{}},\ and\ \citenamefont {Kresse}}]{liu_cubic_gw}%
  \BibitemOpen
  \bibfield  {author} {\bibinfo {author} {\bibfnamefont {P.}~\bibnamefont {Liu}}, \bibinfo {author} {\bibfnamefont {M.}~\bibnamefont {Kaltak}}, \bibinfo {author} {\bibfnamefont {J.~c.~v.}\ \bibnamefont {Klime\ifmmode~\check{s}\else \v{s}\fi{}}}, \ and\ \bibinfo {author} {\bibfnamefont {G.}~\bibnamefont {Kresse}},\ }\href {\doibase 10.1103/PhysRevB.94.165109} {\bibfield  {journal} {\bibinfo  {journal} {Phys. Rev. B}\ }\textbf {\bibinfo {volume} {94}},\ \bibinfo {pages} {165109} (\bibinfo {year} {2016})}\BibitemShut {NoStop}%
\bibitem [{\citenamefont {Duchemin}\ and\ \citenamefont {Blase}(2021)}]{Duchemin2021-cv}%
  \BibitemOpen
  \bibfield  {author} {\bibinfo {author} {\bibfnamefont {I.}~\bibnamefont {Duchemin}}\ and\ \bibinfo {author} {\bibfnamefont {X.}~\bibnamefont {Blase}},\ }\href {\doibase 10.1021/acs.jctc.1c00101} {\bibfield  {journal} {\bibinfo  {journal} {J. Chem. Theory Comput.}\ }\textbf {\bibinfo {volume} {17}},\ \bibinfo {pages} {2383} (\bibinfo {year} {2021})}\BibitemShut {NoStop}%
\bibitem [{\citenamefont {Neuhauser}\ \emph {et~al.}(2014)\citenamefont {Neuhauser}, \citenamefont {Gao}, \citenamefont {Arntsen}, \citenamefont {Karshenas}, \citenamefont {Rabani},\ and\ \citenamefont {Baer}}]{Neuhauser2014}%
  \BibitemOpen
  \bibfield  {author} {\bibinfo {author} {\bibfnamefont {D.}~\bibnamefont {Neuhauser}}, \bibinfo {author} {\bibfnamefont {Y.}~\bibnamefont {Gao}}, \bibinfo {author} {\bibfnamefont {C.}~\bibnamefont {Arntsen}}, \bibinfo {author} {\bibfnamefont {C.}~\bibnamefont {Karshenas}}, \bibinfo {author} {\bibfnamefont {E.}~\bibnamefont {Rabani}}, \ and\ \bibinfo {author} {\bibfnamefont {R.}~\bibnamefont {Baer}},\ }\href {\doibase 10.1103/PhysRevLett.113.076402} {\bibfield  {journal} {\bibinfo  {journal} {Phys. Rev. Lett.}\ }\textbf {\bibinfo {volume} {113}},\ \bibinfo {pages} {076402} (\bibinfo {year} {2014})}\BibitemShut {NoStop}%
\bibitem [{\citenamefont {Jiang}\ \emph {et~al.}(2013)\citenamefont {Jiang}, \citenamefont {Gómez-Abal}, \citenamefont {Li}, \citenamefont {Meisenbichler}, \citenamefont {Ambrosch-Draxl},\ and\ \citenamefont {Scheffler}}]{Jiang2013}%
  \BibitemOpen
  \bibfield  {author} {\bibinfo {author} {\bibfnamefont {H.}~\bibnamefont {Jiang}}, \bibinfo {author} {\bibfnamefont {R.~I.}\ \bibnamefont {Gómez-Abal}}, \bibinfo {author} {\bibfnamefont {X.~Z.}\ \bibnamefont {Li}}, \bibinfo {author} {\bibfnamefont {C.}~\bibnamefont {Meisenbichler}}, \bibinfo {author} {\bibfnamefont {C.}~\bibnamefont {Ambrosch-Draxl}}, \ and\ \bibinfo {author} {\bibfnamefont {M.}~\bibnamefont {Scheffler}},\ }\href {\doibase 10.1016/j.cpc.2012.09.018} {\bibfield  {journal} {\bibinfo  {journal} {Comput.~Phys.~Commun.~}\ }\textbf {\bibinfo {volume} {184}},\ \bibinfo {pages} {348} (\bibinfo {year} {2013})}\BibitemShut {NoStop}%
\bibitem [{\citenamefont {Nabok}\ \emph {et~al.}(2016)\citenamefont {Nabok}, \citenamefont {Gulans},\ and\ \citenamefont {Draxl}}]{Nabok2016}%
  \BibitemOpen
  \bibfield  {author} {\bibinfo {author} {\bibfnamefont {D.}~\bibnamefont {Nabok}}, \bibinfo {author} {\bibfnamefont {A.}~\bibnamefont {Gulans}}, \ and\ \bibinfo {author} {\bibfnamefont {C.}~\bibnamefont {Draxl}},\ }\href {\doibase 10.1103/PhysRevB.94.035118} {\bibfield  {journal} {\bibinfo  {journal} {Phys.~Rev.~B}\ }\textbf {\bibinfo {volume} {94}},\ \bibinfo {pages} {035118} (\bibinfo {year} {2016})}\BibitemShut {NoStop}%
\bibitem [{\citenamefont {Ren}\ \emph {et~al.}(2012)\citenamefont {Ren}, \citenamefont {Rinke}, \citenamefont {Blum}, \citenamefont {Wieferink}, \citenamefont {Tkatchenko}, \citenamefont {Sanfilippo}, \citenamefont {Reuter},\ and\ \citenamefont {Scheffler}}]{Ren2012}%
  \BibitemOpen
  \bibfield  {author} {\bibinfo {author} {\bibfnamefont {X.}~\bibnamefont {Ren}}, \bibinfo {author} {\bibfnamefont {P.}~\bibnamefont {Rinke}}, \bibinfo {author} {\bibfnamefont {V.}~\bibnamefont {Blum}}, \bibinfo {author} {\bibfnamefont {J.}~\bibnamefont {Wieferink}}, \bibinfo {author} {\bibfnamefont {A.}~\bibnamefont {Tkatchenko}}, \bibinfo {author} {\bibfnamefont {A.}~\bibnamefont {Sanfilippo}}, \bibinfo {author} {\bibfnamefont {K.}~\bibnamefont {Reuter}}, \ and\ \bibinfo {author} {\bibfnamefont {M.}~\bibnamefont {Scheffler}},\ }\href {\doibase 10.1088/1367-2630/14/5/053020} {\bibfield  {journal} {\bibinfo  {journal} {New Journal of Physics}\ }\textbf {\bibinfo {volume} {14}},\ \bibinfo {pages} {053020} (\bibinfo {year} {2012})}\BibitemShut {NoStop}%
\bibitem [{\citenamefont {Lejaeghere}\ \emph {et~al.}(2016)\citenamefont {Lejaeghere}, \citenamefont {Bihlmayer}, \citenamefont {Björkman}, \citenamefont {Blaha}, \citenamefont {Blügel}, \citenamefont {Blum}, \citenamefont {Caliste}, \citenamefont {Castelli}, \citenamefont {Clark}, \citenamefont {Corso}, \citenamefont {de~Gironcoli}, \citenamefont {Deutsch}, \citenamefont {Dewhurst}, \citenamefont {Marco}, \citenamefont {Draxl}, \citenamefont {Dułak}, \citenamefont {Eriksson}, \citenamefont {Flores-Livas}, \citenamefont {Garrity}, \citenamefont {Genovese}, \citenamefont {Giannozzi}, \citenamefont {Giantomassi}, \citenamefont {Goedecker}, \citenamefont {Gonze}, \citenamefont {Grånäs}, \citenamefont {Gross}, \citenamefont {Gulans}, \citenamefont {Gygi}, \citenamefont {Hamann}, \citenamefont {Hasnip}, \citenamefont {Holzwarth}, \citenamefont {Iuşan}, \citenamefont {Jochym}, \citenamefont {Jollet}, \citenamefont {Jones}, \citenamefont {Kresse}, \citenamefont {Koepernik}, \citenamefont {Küçükbenli},
  \citenamefont {Kvashnin}, \citenamefont {Locht}, \citenamefont {Lubeck}, \citenamefont {Marsman}, \citenamefont {Marzari}, \citenamefont {Nitzsche}, \citenamefont {Nordström}, \citenamefont {Ozaki}, \citenamefont {Paulatto}, \citenamefont {Pickard}, \citenamefont {Poelmans}, \citenamefont {Probert}, \citenamefont {Refson}, \citenamefont {Richter}, \citenamefont {Rignanese}, \citenamefont {Saha}, \citenamefont {Scheffler}, \citenamefont {Schlipf}, \citenamefont {Schwarz}, \citenamefont {Sharma}, \citenamefont {Tavazza}, \citenamefont {Thunström}, \citenamefont {Tkatchenko}, \citenamefont {Torrent}, \citenamefont {Vanderbilt}, \citenamefont {van Setten}, \citenamefont {Speybroeck}, \citenamefont {Wills}, \citenamefont {Yates}, \citenamefont {Zhang},\ and\ \citenamefont {Cottenier}}]{Lejaeghere2016}%
  \BibitemOpen
  \bibfield  {author} {\bibinfo {author} {\bibfnamefont {K.}~\bibnamefont {Lejaeghere}}, \bibinfo {author} {\bibfnamefont {G.}~\bibnamefont {Bihlmayer}}, \bibinfo {author} {\bibfnamefont {T.}~\bibnamefont {Björkman}}, \bibinfo {author} {\bibfnamefont {P.}~\bibnamefont {Blaha}}, \bibinfo {author} {\bibfnamefont {S.}~\bibnamefont {Blügel}}, \bibinfo {author} {\bibfnamefont {V.}~\bibnamefont {Blum}}, \bibinfo {author} {\bibfnamefont {D.}~\bibnamefont {Caliste}}, \bibinfo {author} {\bibfnamefont {I.~E.}\ \bibnamefont {Castelli}}, \bibinfo {author} {\bibfnamefont {S.~J.}\ \bibnamefont {Clark}}, \bibinfo {author} {\bibfnamefont {A.~D.}\ \bibnamefont {Corso}}, \bibinfo {author} {\bibfnamefont {S.}~\bibnamefont {de~Gironcoli}}, \bibinfo {author} {\bibfnamefont {T.}~\bibnamefont {Deutsch}}, \bibinfo {author} {\bibfnamefont {J.~K.}\ \bibnamefont {Dewhurst}}, \bibinfo {author} {\bibfnamefont {I.~D.}\ \bibnamefont {Marco}}, \bibinfo {author} {\bibfnamefont {C.}~\bibnamefont {Draxl}}, \bibinfo {author} {\bibfnamefont
  {M.}~\bibnamefont {Dułak}}, \bibinfo {author} {\bibfnamefont {O.}~\bibnamefont {Eriksson}}, \bibinfo {author} {\bibfnamefont {J.~A.}\ \bibnamefont {Flores-Livas}}, \bibinfo {author} {\bibfnamefont {K.~F.}\ \bibnamefont {Garrity}}, \bibinfo {author} {\bibfnamefont {L.}~\bibnamefont {Genovese}}, \bibinfo {author} {\bibfnamefont {P.}~\bibnamefont {Giannozzi}}, \bibinfo {author} {\bibfnamefont {M.}~\bibnamefont {Giantomassi}}, \bibinfo {author} {\bibfnamefont {S.}~\bibnamefont {Goedecker}}, \bibinfo {author} {\bibfnamefont {X.}~\bibnamefont {Gonze}}, \bibinfo {author} {\bibfnamefont {O.}~\bibnamefont {Grånäs}}, \bibinfo {author} {\bibfnamefont {E.~K.~U.}\ \bibnamefont {Gross}}, \bibinfo {author} {\bibfnamefont {A.}~\bibnamefont {Gulans}}, \bibinfo {author} {\bibfnamefont {F.}~\bibnamefont {Gygi}}, \bibinfo {author} {\bibfnamefont {D.~R.}\ \bibnamefont {Hamann}}, \bibinfo {author} {\bibfnamefont {P.~J.}\ \bibnamefont {Hasnip}}, \bibinfo {author} {\bibfnamefont {N.~A.~W.}\ \bibnamefont {Holzwarth}}, \bibinfo
  {author} {\bibfnamefont {D.}~\bibnamefont {Iuşan}}, \bibinfo {author} {\bibfnamefont {D.~B.}\ \bibnamefont {Jochym}}, \bibinfo {author} {\bibfnamefont {F.}~\bibnamefont {Jollet}}, \bibinfo {author} {\bibfnamefont {D.}~\bibnamefont {Jones}}, \bibinfo {author} {\bibfnamefont {G.}~\bibnamefont {Kresse}}, \bibinfo {author} {\bibfnamefont {K.}~\bibnamefont {Koepernik}}, \bibinfo {author} {\bibfnamefont {E.}~\bibnamefont {Küçükbenli}}, \bibinfo {author} {\bibfnamefont {Y.~O.}\ \bibnamefont {Kvashnin}}, \bibinfo {author} {\bibfnamefont {I.~L.~M.}\ \bibnamefont {Locht}}, \bibinfo {author} {\bibfnamefont {S.}~\bibnamefont {Lubeck}}, \bibinfo {author} {\bibfnamefont {M.}~\bibnamefont {Marsman}}, \bibinfo {author} {\bibfnamefont {N.}~\bibnamefont {Marzari}}, \bibinfo {author} {\bibfnamefont {U.}~\bibnamefont {Nitzsche}}, \bibinfo {author} {\bibfnamefont {L.}~\bibnamefont {Nordström}}, \bibinfo {author} {\bibfnamefont {T.}~\bibnamefont {Ozaki}}, \bibinfo {author} {\bibfnamefont {L.}~\bibnamefont {Paulatto}},
  \bibinfo {author} {\bibfnamefont {C.~J.}\ \bibnamefont {Pickard}}, \bibinfo {author} {\bibfnamefont {W.}~\bibnamefont {Poelmans}}, \bibinfo {author} {\bibfnamefont {M.~I.~J.}\ \bibnamefont {Probert}}, \bibinfo {author} {\bibfnamefont {K.}~\bibnamefont {Refson}}, \bibinfo {author} {\bibfnamefont {M.}~\bibnamefont {Richter}}, \bibinfo {author} {\bibfnamefont {G.-M.}\ \bibnamefont {Rignanese}}, \bibinfo {author} {\bibfnamefont {S.}~\bibnamefont {Saha}}, \bibinfo {author} {\bibfnamefont {M.}~\bibnamefont {Scheffler}}, \bibinfo {author} {\bibfnamefont {M.}~\bibnamefont {Schlipf}}, \bibinfo {author} {\bibfnamefont {K.}~\bibnamefont {Schwarz}}, \bibinfo {author} {\bibfnamefont {S.}~\bibnamefont {Sharma}}, \bibinfo {author} {\bibfnamefont {F.}~\bibnamefont {Tavazza}}, \bibinfo {author} {\bibfnamefont {P.}~\bibnamefont {Thunström}}, \bibinfo {author} {\bibfnamefont {A.}~\bibnamefont {Tkatchenko}}, \bibinfo {author} {\bibfnamefont {M.}~\bibnamefont {Torrent}}, \bibinfo {author} {\bibfnamefont {D.}~\bibnamefont
  {Vanderbilt}}, \bibinfo {author} {\bibfnamefont {M.~J.}\ \bibnamefont {van Setten}}, \bibinfo {author} {\bibfnamefont {V.~V.}\ \bibnamefont {Speybroeck}}, \bibinfo {author} {\bibfnamefont {J.~M.}\ \bibnamefont {Wills}}, \bibinfo {author} {\bibfnamefont {J.~R.}\ \bibnamefont {Yates}}, \bibinfo {author} {\bibfnamefont {G.-X.}\ \bibnamefont {Zhang}}, \ and\ \bibinfo {author} {\bibfnamefont {S.}~\bibnamefont {Cottenier}},\ }\href {\doibase 10.1126/science.aad3000} {\bibfield  {journal} {\bibinfo  {journal} {Science}\ }\textbf {\bibinfo {volume} {351}},\ \bibinfo {pages} {aad3000} (\bibinfo {year} {2016})}\BibitemShut {NoStop}%
\bibitem [{\citenamefont {Azizi}\ \emph {et~al.}(2025)\citenamefont {Azizi}, \citenamefont {Delesma}, \citenamefont {Giantomassi}, \citenamefont {Zavickis}, \citenamefont {Kuisma}, \citenamefont {Thyghesen}, \citenamefont {Golze}, \citenamefont {Buccheri}, \citenamefont {Zhang}, \citenamefont {Rinke}, \citenamefont {Draxl}, \citenamefont {Gulans},\ and\ \citenamefont {Gonze}}]{Azizi2025}%
  \BibitemOpen
  \bibfield  {author} {\bibinfo {author} {\bibfnamefont {M.}~\bibnamefont {Azizi}}, \bibinfo {author} {\bibfnamefont {F.~A.}\ \bibnamefont {Delesma}}, \bibinfo {author} {\bibfnamefont {M.}~\bibnamefont {Giantomassi}}, \bibinfo {author} {\bibfnamefont {D.}~\bibnamefont {Zavickis}}, \bibinfo {author} {\bibfnamefont {M.}~\bibnamefont {Kuisma}}, \bibinfo {author} {\bibfnamefont {K.}~\bibnamefont {Thyghesen}}, \bibinfo {author} {\bibfnamefont {D.}~\bibnamefont {Golze}}, \bibinfo {author} {\bibfnamefont {A.}~\bibnamefont {Buccheri}}, \bibinfo {author} {\bibfnamefont {M.-Y.}\ \bibnamefont {Zhang}}, \bibinfo {author} {\bibfnamefont {P.}~\bibnamefont {Rinke}}, \bibinfo {author} {\bibfnamefont {C.}~\bibnamefont {Draxl}}, \bibinfo {author} {\bibfnamefont {A.}~\bibnamefont {Gulans}}, \ and\ \bibinfo {author} {\bibfnamefont {X.}~\bibnamefont {Gonze}},\ }\href {\doibase https://doi.org/10.1016/j.commatsci.2024.113655} {\bibfield  {journal} {\bibinfo  {journal} {Computational Materials Science}\ }\textbf {\bibinfo {volume}
  {250}},\ \bibinfo {pages} {113655} (\bibinfo {year} {2025})}\BibitemShut {NoStop}%
\bibitem [{\citenamefont {Singh}\ and\ \citenamefont {Nordström}(2005)}]{Singh2005}%
  \BibitemOpen
  \bibfield  {author} {\bibinfo {author} {\bibfnamefont {D.~J.}\ \bibnamefont {Singh}}\ and\ \bibinfo {author} {\bibfnamefont {L.}~\bibnamefont {Nordström}},\ }\href {\doibase https://doi.org/10.1007/978-0-387-29684-5} {\emph {\bibinfo {title} {{Planewaves, Pseudopotentials, and the LAPW Method}}}}\ (\bibinfo  {publisher} {Springer New York, NY},\ \bibinfo {year} {2005})\BibitemShut {NoStop}%
\bibitem [{\citenamefont {Gulans}\ \emph {et~al.}(2014)\citenamefont {Gulans}, \citenamefont {Kontur}, \citenamefont {Meisenbichler}, \citenamefont {Nabok}, \citenamefont {Pavone}, \citenamefont {Rigamonti}, \citenamefont {Sagmeister}, \citenamefont {Werner},\ and\ \citenamefont {Draxl}}]{Gulans2014}%
  \BibitemOpen
  \bibfield  {author} {\bibinfo {author} {\bibfnamefont {A.}~\bibnamefont {Gulans}}, \bibinfo {author} {\bibfnamefont {S.}~\bibnamefont {Kontur}}, \bibinfo {author} {\bibfnamefont {C.}~\bibnamefont {Meisenbichler}}, \bibinfo {author} {\bibfnamefont {D.}~\bibnamefont {Nabok}}, \bibinfo {author} {\bibfnamefont {P.}~\bibnamefont {Pavone}}, \bibinfo {author} {\bibfnamefont {S.}~\bibnamefont {Rigamonti}}, \bibinfo {author} {\bibfnamefont {S.}~\bibnamefont {Sagmeister}}, \bibinfo {author} {\bibfnamefont {U.}~\bibnamefont {Werner}}, \ and\ \bibinfo {author} {\bibfnamefont {C.}~\bibnamefont {Draxl}},\ }\href {\doibase 10.1088/0953-8984/26/36/363202} {\bibfield  {journal} {\bibinfo  {journal} {J.~Phys.~Condens.~Matter.~}\ }\textbf {\bibinfo {volume} {26}},\ \bibinfo {pages} {363202} (\bibinfo {year} {2014})}\BibitemShut {NoStop}%
\bibitem [{\citenamefont {Vorwerk}\ \emph {et~al.}(2019)\citenamefont {Vorwerk}, \citenamefont {Aurich}, \citenamefont {Cocchi},\ and\ \citenamefont {Draxl}}]{Vorwerk2019}%
  \BibitemOpen
  \bibfield  {author} {\bibinfo {author} {\bibfnamefont {C.}~\bibnamefont {Vorwerk}}, \bibinfo {author} {\bibfnamefont {B.}~\bibnamefont {Aurich}}, \bibinfo {author} {\bibfnamefont {C.}~\bibnamefont {Cocchi}}, \ and\ \bibinfo {author} {\bibfnamefont {C.}~\bibnamefont {Draxl}},\ }\href {\doibase 10.1088/2516-1075/ab3123} {\bibfield  {journal} {\bibinfo  {journal} {Electron.~Struct.~}\ }\textbf {\bibinfo {volume} {1}},\ \bibinfo {pages} {037001} (\bibinfo {year} {2019})}\BibitemShut {NoStop}%
\bibitem [{\citenamefont {Rohlfing}\ and\ \citenamefont {Louie}(2000)}]{Rohlfing2000}%
  \BibitemOpen
  \bibfield  {author} {\bibinfo {author} {\bibfnamefont {M.}~\bibnamefont {Rohlfing}}\ and\ \bibinfo {author} {\bibfnamefont {S.~G.}\ \bibnamefont {Louie}},\ }\href {https://journals.aps.org/prb/abstract/10.1103/PhysRevB.62.4927} {\bibfield  {journal} {\bibinfo  {journal} {Phys.~Rev.~B}\ }\textbf {\bibinfo {volume} {62}},\ \bibinfo {pages} {4927} (\bibinfo {year} {2000})}\BibitemShut {NoStop}%
\bibitem [{\citenamefont {Friedrich}\ \emph {et~al.}(2010)\citenamefont {Friedrich}, \citenamefont {Bl\"ugel},\ and\ \citenamefont {Schindlmayr}}]{Friedrich2010}%
  \BibitemOpen
  \bibfield  {author} {\bibinfo {author} {\bibfnamefont {C.}~\bibnamefont {Friedrich}}, \bibinfo {author} {\bibfnamefont {S.}~\bibnamefont {Bl\"ugel}}, \ and\ \bibinfo {author} {\bibfnamefont {A.}~\bibnamefont {Schindlmayr}},\ }\href {\doibase 10.1103/PhysRevB.81.125102} {\bibfield  {journal} {\bibinfo  {journal} {Phys.~Rev.~B}\ }\textbf {\bibinfo {volume} {81}},\ \bibinfo {pages} {125102} (\bibinfo {year} {2010})}\BibitemShut {NoStop}%
\bibitem [{\citenamefont {He}\ \emph {et~al.}(2014)\citenamefont {He}, \citenamefont {Hummer},\ and\ \citenamefont {Franchini}}]{He2014}%
  \BibitemOpen
  \bibfield  {author} {\bibinfo {author} {\bibfnamefont {J.}~\bibnamefont {He}}, \bibinfo {author} {\bibfnamefont {K.}~\bibnamefont {Hummer}}, \ and\ \bibinfo {author} {\bibfnamefont {C.}~\bibnamefont {Franchini}},\ }\href {\doibase 10.1103/PhysRevB.89.075409} {\bibfield  {journal} {\bibinfo  {journal} {Phys.~Rev.~B}\ }\textbf {\bibinfo {volume} {89}},\ \bibinfo {pages} {075409} (\bibinfo {year} {2014})}\BibitemShut {NoStop}%
\bibitem [{\citenamefont {Tkatchenko}\ and\ \citenamefont {Scheffler}(2009)}]{Tkatchenko2009}%
  \BibitemOpen
  \bibfield  {author} {\bibinfo {author} {\bibfnamefont {A.}~\bibnamefont {Tkatchenko}}\ and\ \bibinfo {author} {\bibfnamefont {M.}~\bibnamefont {Scheffler}},\ }\href {\doibase 10.1103/PhysRevLett.102.073005} {\bibfield  {journal} {\bibinfo  {journal} {Phys.~Rev.~Lett.~}\ }\textbf {\bibinfo {volume} {102}},\ \bibinfo {pages} {073005} (\bibinfo {year} {2009})}\BibitemShut {NoStop}%
\bibitem [{\citenamefont {Tkatchenko}\ \emph {et~al.}(2010)\citenamefont {Tkatchenko}, \citenamefont {Romaner}, \citenamefont {Hofmann}, \citenamefont {Zojer}, \citenamefont {Ambrosch-Draxl},\ and\ \citenamefont {Scheffler}}]{Tkatchenko2010}%
  \BibitemOpen
  \bibfield  {author} {\bibinfo {author} {\bibfnamefont {A.}~\bibnamefont {Tkatchenko}}, \bibinfo {author} {\bibfnamefont {L.}~\bibnamefont {Romaner}}, \bibinfo {author} {\bibfnamefont {O.~T.}\ \bibnamefont {Hofmann}}, \bibinfo {author} {\bibfnamefont {E.}~\bibnamefont {Zojer}}, \bibinfo {author} {\bibfnamefont {C.}~\bibnamefont {Ambrosch-Draxl}}, \ and\ \bibinfo {author} {\bibfnamefont {M.}~\bibnamefont {Scheffler}},\ }\href {\doibase 10.1557/mrs2010.581} {\bibfield  {journal} {\bibinfo  {journal} {MRS Bulletin}\ }\textbf {\bibinfo {volume} {35}},\ \bibinfo {pages} {435} (\bibinfo {year} {2010})}\BibitemShut {NoStop}%
\bibitem [{\citenamefont {Rodrigues~Pela}\ \emph {et~al.}(2024)\citenamefont {Rodrigues~Pela}, \citenamefont {Vona}, \citenamefont {Lubeck}, \citenamefont {Alex}, \citenamefont {Gonzalez~Oliva},\ and\ \citenamefont {Draxl}}]{RodriguesPela2024}%
  \BibitemOpen
  \bibfield  {author} {\bibinfo {author} {\bibfnamefont {R.}~\bibnamefont {Rodrigues~Pela}}, \bibinfo {author} {\bibfnamefont {C.}~\bibnamefont {Vona}}, \bibinfo {author} {\bibfnamefont {S.}~\bibnamefont {Lubeck}}, \bibinfo {author} {\bibfnamefont {B.}~\bibnamefont {Alex}}, \bibinfo {author} {\bibfnamefont {I.}~\bibnamefont {Gonzalez~Oliva}}, \ and\ \bibinfo {author} {\bibfnamefont {C.}~\bibnamefont {Draxl}},\ }\href {\doibase 10.1038/s41524-024-01253-2} {\bibfield  {journal} {\bibinfo  {journal} {npj Comput. Mater.}\ }\textbf {\bibinfo {volume} {10}},\ \bibinfo {pages} {77} (\bibinfo {year} {2024})}\BibitemShut {NoStop}%
\bibitem [{\citenamefont {Haastrup}\ \emph {et~al.}(2018)\citenamefont {Haastrup}, \citenamefont {Strange}, \citenamefont {Pandey}, \citenamefont {Deilmann}, \citenamefont {Schmidt}, \citenamefont {Hinsche}, \citenamefont {Gjerding}, \citenamefont {Torelli}, \citenamefont {Larsen}, \citenamefont {Riis-Jensen}, \citenamefont {Gath}, \citenamefont {Jacobsen}, \citenamefont {Mortensen}, \citenamefont {Olsen},\ and\ \citenamefont {Thygesen}}]{Haastrup_2018}%
  \BibitemOpen
  \bibfield  {author} {\bibinfo {author} {\bibfnamefont {S.}~\bibnamefont {Haastrup}}, \bibinfo {author} {\bibfnamefont {M.}~\bibnamefont {Strange}}, \bibinfo {author} {\bibfnamefont {M.}~\bibnamefont {Pandey}}, \bibinfo {author} {\bibfnamefont {T.}~\bibnamefont {Deilmann}}, \bibinfo {author} {\bibfnamefont {P.~S.}\ \bibnamefont {Schmidt}}, \bibinfo {author} {\bibfnamefont {N.~F.}\ \bibnamefont {Hinsche}}, \bibinfo {author} {\bibfnamefont {M.~N.}\ \bibnamefont {Gjerding}}, \bibinfo {author} {\bibfnamefont {D.}~\bibnamefont {Torelli}}, \bibinfo {author} {\bibfnamefont {P.~M.}\ \bibnamefont {Larsen}}, \bibinfo {author} {\bibfnamefont {A.~C.}\ \bibnamefont {Riis-Jensen}}, \bibinfo {author} {\bibfnamefont {J.}~\bibnamefont {Gath}}, \bibinfo {author} {\bibfnamefont {K.~W.}\ \bibnamefont {Jacobsen}}, \bibinfo {author} {\bibfnamefont {J.~J.}\ \bibnamefont {Mortensen}}, \bibinfo {author} {\bibfnamefont {T.}~\bibnamefont {Olsen}}, \ and\ \bibinfo {author} {\bibfnamefont {K.~S.}\ \bibnamefont {Thygesen}},\ }\href
  {\doibase 10.1088/2053-1583/aacfc1} {\bibfield  {journal} {\bibinfo  {journal} {2D Materials}\ }\textbf {\bibinfo {volume} {5}},\ \bibinfo {pages} {042002} (\bibinfo {year} {2018})}\BibitemShut {NoStop}%
\bibitem [{\citenamefont {Gjerding}\ \emph {et~al.}(2021)\citenamefont {Gjerding}, \citenamefont {Taghizadeh}, \citenamefont {Rasmussen}, \citenamefont {Ali}, \citenamefont {Bertoldo}, \citenamefont {Deilmann}, \citenamefont {Knøsgaard}, \citenamefont {Kruse}, \citenamefont {Larsen}, \citenamefont {Manti}, \citenamefont {Pedersen}, \citenamefont {Petralanda}, \citenamefont {Skovhus}, \citenamefont {Svendsen}, \citenamefont {Mortensen}, \citenamefont {Olsen},\ and\ \citenamefont {Thygesen}}]{Gjerding_2021}%
  \BibitemOpen
  \bibfield  {author} {\bibinfo {author} {\bibfnamefont {M.~N.}\ \bibnamefont {Gjerding}}, \bibinfo {author} {\bibfnamefont {A.}~\bibnamefont {Taghizadeh}}, \bibinfo {author} {\bibfnamefont {A.}~\bibnamefont {Rasmussen}}, \bibinfo {author} {\bibfnamefont {S.}~\bibnamefont {Ali}}, \bibinfo {author} {\bibfnamefont {F.}~\bibnamefont {Bertoldo}}, \bibinfo {author} {\bibfnamefont {T.}~\bibnamefont {Deilmann}}, \bibinfo {author} {\bibfnamefont {N.~R.}\ \bibnamefont {Knøsgaard}}, \bibinfo {author} {\bibfnamefont {M.}~\bibnamefont {Kruse}}, \bibinfo {author} {\bibfnamefont {A.~H.}\ \bibnamefont {Larsen}}, \bibinfo {author} {\bibfnamefont {S.}~\bibnamefont {Manti}}, \bibinfo {author} {\bibfnamefont {T.~G.}\ \bibnamefont {Pedersen}}, \bibinfo {author} {\bibfnamefont {U.}~\bibnamefont {Petralanda}}, \bibinfo {author} {\bibfnamefont {T.}~\bibnamefont {Skovhus}}, \bibinfo {author} {\bibfnamefont {M.~K.}\ \bibnamefont {Svendsen}}, \bibinfo {author} {\bibfnamefont {J.~J.}\ \bibnamefont {Mortensen}}, \bibinfo {author}
  {\bibfnamefont {T.}~\bibnamefont {Olsen}}, \ and\ \bibinfo {author} {\bibfnamefont {K.~S.}\ \bibnamefont {Thygesen}},\ }\href {\doibase 10.1088/2053-1583/ac1059} {\bibfield  {journal} {\bibinfo  {journal} {2D Materials}\ }\textbf {\bibinfo {volume} {8}},\ \bibinfo {pages} {044002} (\bibinfo {year} {2021})}\BibitemShut {NoStop}%
\bibitem [{Note1()}]{Note1}%
  \BibitemOpen
  \bibinfo {note} {CPU hours are defined as the number of processors used multiplied by the number of wall-time hours needed for the calculation.}\BibitemShut {Stop}%
\bibitem [{\citenamefont {Henneke}\ \emph {et~al.}(2020)\citenamefont {Henneke}, \citenamefont {Lin}, \citenamefont {Vorwerk}, \citenamefont {Draxl}, \citenamefont {Klein},\ and\ \citenamefont {Yang}}]{Henneke2020}%
  \BibitemOpen
  \bibfield  {author} {\bibinfo {author} {\bibfnamefont {F.}~\bibnamefont {Henneke}}, \bibinfo {author} {\bibfnamefont {L.}~\bibnamefont {Lin}}, \bibinfo {author} {\bibfnamefont {C.}~\bibnamefont {Vorwerk}}, \bibinfo {author} {\bibfnamefont {C.}~\bibnamefont {Draxl}}, \bibinfo {author} {\bibfnamefont {R.}~\bibnamefont {Klein}}, \ and\ \bibinfo {author} {\bibfnamefont {C.}~\bibnamefont {Yang}},\ }\href {\doibase 10.2140/camcos.2020.15.89} {\bibfield  {journal} {\bibinfo  {journal} {Comm. App. Math. Comp. Sci.}\ }\textbf {\bibinfo {volume} {15}},\ \bibinfo {pages} {89} (\bibinfo {year} {2020})}\BibitemShut {NoStop}%
\end{thebibliography}%

\end{document}